\documentclass[11pt]{article}
\usepackage{epsfig}

\newcommand{\be}{\begin{eqnarray}}
\newcommand{\ee}{\end{eqnarray}}

\def\ll#1{\left#1}
\def\r#1{\right#1}
\def\fr{\frac{1}{2}}

\def\mref#1{(\ref{#1})}

\def\p{\partial}

\def\bd{\begin{displaymath}}
\def\ed{\end{displaymath}}

\def\ba#1{\begin{array}{#1}}
\def\ea{\end{array}}
\def\nn{\nonumber}
\newfont{\Bbb}{msbm10 scaled 1200}

\newtheorem{tw}{THEOREM}
\newtheorem{de}{DEFINITION}

\begin{document}

\pagestyle{empty}

\begin{center}

{\LARGE\bf Semiclassical wave functions and energy spectra in polygon billiards\\[0.5cm]}

\vskip 12pt

{\large {\bf Stefan Giller}}

\vskip 3pt

Jan D{\l}ugosz University in Czestochowa\\ Institute of Physics\\ Armii Krajowej 13/15, 42-200 Czestochowa,
Poland\\ e-mail: stefan.giller@ajd.czest.pl
\end{center}
\vspace{6pt}
\begin{abstract}
A consistent scheme of semiclassical quantization in polygon billiards by wave function formalism is presented. It
is argued that it is in the spirit of the semiclassical wave function formalism to make necessary rationalization of
respective
quantities accompanied the procedure of the semiclassical quantization in polygon billiards. Unfolding rational polygon billiards (RPB)
into corresponding Riemann surfaces (RS) periodic structures of the latter are demonstrated with $2g$
independent periods on the respective multitori with $g$ as their genuses. However it is the two dimensional real space of the {\it real} linear
combinations of these periods which is just used for quantizing RPB's. Next a class of doubly rational polygon billiards (DRPB) is considered
for which these real linear relations are {\it rational} and their semiclassical quantization by wave function
formalism is presented.
It is then shown that semiclassical quantization of both the classical momenta
and the energy spectra are determined completely by periodic structure of the corresponding RS's. Each RS
can be then reduced to elementary polygon patterns (EPP) as its basic periodic elements building it. Each such
EPP can be glued to a torus of genus $g$. Semiclassical wave functions (SWF) are then constructed on EPP's. The SWF's for DRPB's
appear to be exact and have forms of coherent sums of plane waves.
They satisfy on the billiards boundaries well defined conditions - the Dirichlet, the Neumannn or the mixed ones. Not every mixing of such conditions
is allowed however and the respective limitations can ignore some semiclassical states in the presented formalism. A respective incompleteness of
SWF's provided by the method used in the paper is discussed. Families of DRPB's can form dens subsets of {\it angle similar rational}
polygon billiards allowing for approximate semiclassical quantization of the latter. Next general rational polygons are
quantized by approximating them by doubly rational ones. A natural extension of the formalism to irrational polygons is described shortly as well.
When the semiclassical approximations constructed in the paper appear really as only approximations the latter are controlled by a general criteria
of the eigenvalue theory. Finally a relation between the superscar solutions and SWF's constructed in the paper is also discussed.
\end{abstract}

\vskip 3pt

\begin{tabular}{l}
{\small PACS number(s): 03.65.-w, 03.65.Sq, 02.30.Jr, 02.30.Lt} \\[1mm] {\small Key Words:
Schr\"{o}dinger equation, semiclassical expansion, Lagrange manifolds, classical}\\[1mm]{\small trajectories,
rational polygon billiards, integrability, pseudointegrability, chaotic dynamics,}\\{\small quantum chaos, superscars}
\end{tabular}

\newpage

\pagestyle{plain}

\setcounter{page}{1}

\section{Introduction}

\hskip+2em It is well known that the classical dynamics in the irrational polygon billiards (IPB) are chaotic while in the
rational polygon billiards (RPB) this dynamics can be considered as medium between the integrable and the chaotic ones being
described as pseudointegrable \cite{8,37,29}. Indeed only a few cases of the RPB are known which are integrable, i.e. the
rectangle and several triangle billiards while the remaining pseudointegrable RPB's are represented in the corresponding
phase space by closed Lagrange surfaces with genuses $g>1$. For the irrational billiards none Lagrange surface
exists at all \cite{5}.

The above circumstances are bases for claiming that classical motions in the irrational billiards cannot be quantized
semiclassically at all by the wave function formulation language as well as in the case of the rational polygon
billiards despite the fact of existence of respective Lagrange surfaces since the
basic quantities defined on the surfaces - the actions - cannot be quantized independently in a consistent way
satisfying geometry of billiards \cite{8}.

Such a point of view can be however criticised having in mind that the semiclassical approach is just an approximation
to the exact wave functions introducing natural length measures - the wave lengths. Because of that the Lagrange
surface periods have all to be measured by the wave length units providing us with integer numbers as results of such
measurements, i.e. all these periods should be commensurate. Since the last situation can be rather exceptional
than typical one has to consider it as approximate i.e. as being satisfied with a sufficient accuracy.

This situation is
in fact similar to the one where we are to compare a side of the square with its diagonal having a particular unit of
measure which provide us with an integer number of the side length say 100. For the diagonal length we get then as we
well know the number 141 plus the rest of its length smaller than 1, i.e. the square side and its diagonal are
commensurate within the accuracy of the used measure unit. We can of course improve this commensurateness to an
arbitrary level of accuracy by diminishing the used measure unit.

Therefore to be consequent in applications of the semiclassical approximations we have to be ready also for accepting
necessary approximations appearing as results of using the wave lengths provided by the semiclassical approach as the
natural units of length measurements. It is clear that an accuracy of such approximations are the better the shorter
are the wave lengths used to measure the respective length quantities. But such a situation is just in an agreement with
the assumed validity of the semiclassical approximation which is to be the better the higher energy are considered, i.e.
the shorter waves dominates in the quantum problem considered.

It is the aim of this paper to describe the way in which RPB's can be
quantized semiclassically according to the "philosophy" described above and in a consistent way.

As main areas on which our goal is realized are Riemann surfaces developed by unfolding rational polygons
considered. Such
surfaces have periodic structures formed by elementary patterns of polygons periodically distributed on the surfaces.
An even number of periods of a RS equal to $2g$ corresponds to a multitorus of genus $g$ obtained by gluing respective boundaries of EPP's.
Semiclassical wave functions are defined on skeletons totally covering RS's on which they have to satisfy periodic conditions with these $2g$
independent periods. The latter conditions can be satisfied however only by SWF's constructed on a special class of RPB's, i.e. on DRPB's
the linear {\it real} relations between their periods on the plane are in fact {\it rational}.
The rational periodic conditions demand then the classical momenta to be quantized. The periodicity of SWF's on DRPB's is also the key
for the final forms of them which are coherent sums of plane waves and determines also their corresponding energy spectra.
The approach applied to DRPB's can be used to quantized semiclassically those families of RPB's which contain the respective DRPB's as their dens subsets.
The approximate description of the energy levels are then controlled by general theorems on the behavior of the levels \cite{40}. A natural extension
of the method can be next done to IPB's being controlled by the general theorem just mentioned \cite{40}.

Our paper is organized as follows.

In sec.2 the main tools of our approach are presented and it is shown how with their help one can quantize the
classical momenta of the billiard ball in any DRPB.

In the next section energy spectrum corresponding to periodic and aperiodic skeletons are established.

In sec.4 SWF's are constructed on periodic and aperiodic skeletons satisfying boundary conditions allowed by polygons
considered.

In sec.5 the procedure of semiclassical quantization developed in the previous sections are applied to the family of parallelogram
billiards with the smaller angle equal to $\pi/3$ and to its broken single bay version and to the family of single bay rectangular billiards.

In sec.6 the generally unavoidable incompleteness of the SWF's generated by the method presented in the paper is discussed.

In sec.7 extensions of our approach to RPB's deprived of dens subsets od DRPB's as well as to IPB are discussed.

In sec.8 periodic orbit channel (POC) skeletons of Bogommolny and Schmit \cite{1,2} building global periodic skeletons are considered and their
relations to these skeletons are discussed.

In sec.9 our results are summarized and discussed.

In app.A a short list of notions and acronyms used in the paper is attached.

In app.B a construction of basic semiclassical wave functions in periodic skeletons containing POC's are discussed.

In app.C the main theorems are cited \cite{40} which describe the behaviors of energy levels when the domain boundaries are varied.

\section{Unfolding rational polygons, skeletons and semiclassical\\ wave functions defined on them}

\hskip+2em The unfolding technique, i.e. subsequent mirror reflections of a ray and the polygon considered by
polygon edges, which substitute a real motion of the ray by its motion along a straight line crossing subsequent
mirror images of the polygon is well known in investigations of the polygon billiard dynamics. In simple cases of
the integrable polygon billiard motions (in several triangle billiards, in the rectangle ones) such unfolding is
simple and the respective unfolded polygons cover simply the plane. For the reminder of the cases their total
unfolding can become extremely complicated even for simply looking billiards.

\subsection{Rational polygon Riemann surface structure and its relation with tori of genus $g$}

\hskip+2em Consider a rational polygon billiard Fig.1 with its angles $\alpha_k$ equal to $\frac{p_k}{q_k}\pi$
where $p_k$ and $q_k,\;k=1,...,n$, are coprime integers.
Unfolding the polygon around the angle $\alpha_k$ by reflecting it subsequently by the two edges making the angle we come back to its initial
position after making $2q_k$ such reflections by which each edge of the angle is turned by $2p_k\pi$ around the
angle vertex. Therefore if $p_k>1$ the polygon is unfolded locally by such a vertex onto $p_k$ planes which locally
have the Riemann surface structure with the vertex as the branch point. Every such a vertex will be called {\it branching}, i.e.
a vertex with $p_k=1$ is {\it no} branching.

If further such unfoldings are made around each polygon vertex we
get a branching figure which consists of $p_k$ planes depending on the $k$-th vertex, $k=1,...,n$
for the $n$-vertex polygon and forming locally the Riemann surface structure. However these local Riemann surface
structures cannot be glued in general into a global one composed of a finite number of planes except a few cases of such
unfoldings one of which is the $\pi/3$-rhombus.

\begin{figure}
\begin{center}
\psfig{figure=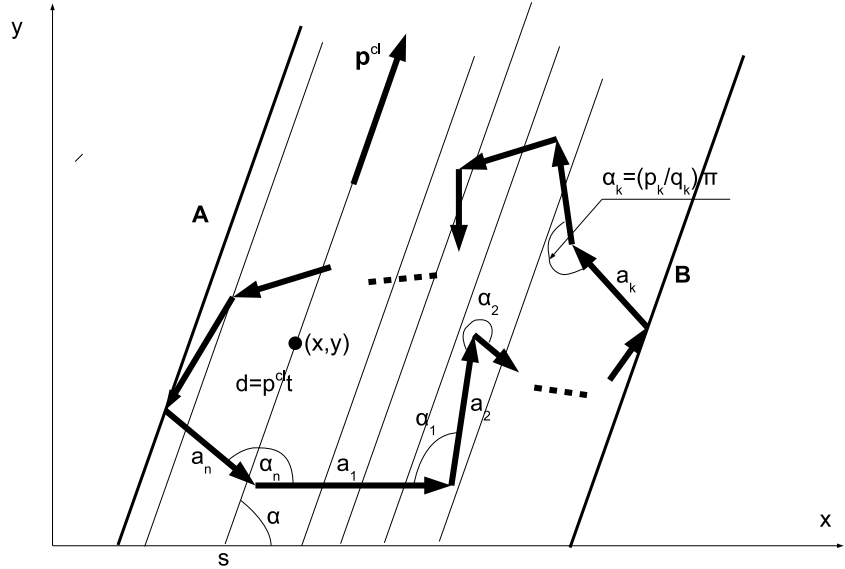,width=12cm} \caption{A rational polygon billiard with a global
skeleton defined by its global compound bundle bounded by the rays {\bf A} and {\bf B} and bearing a classical momentum
${\bf p}^{cl}$}
\end{center}
\end{figure}

Therefore a figure provided by unfolding a rational polygon is in general infinitely branching with
an infinite number of branching points determining only a local structure of the figure. Its global form will be called
the rational polygon Riemann surface (RPRS). The complexity of RPRS is the reason
why trying to unfold a rational polygon on the plane one find such a task almost impossible to be done even for simpler
cases of such polygons.

However each particular rational polygon developes its unique RS. When making it the original polygon changes its orientation
after each mirror reflection so that after each two subsequent reflections it is rotated by an angle defined by the
edges by which it is reflected. However after a finite even number of such reflections the rational polygon
always comes to a position to which it can be brought by a finite translation from its original one. In each such a
position
a polygon is a {\it faithful} copy of the original one. Each such a copy will be therefore called {\it faithful} while the
original polygon will be called {\it basic} (BRPB). The corresponding translation is a {\it period} of RPRS which of course is
one of infinitely many such periods of
RPRS since there are infinitely many of faithful positions periodically repeated on RPRS.

Note however that any image polygon can be chosen as basic and a RPRS generated by such a choice is the same, i.e. the
RPRS as well as its periodic structure is invariant on a choice of a basic polygon which means also that the RPRS is
invariant under action of any of its periods mentioned above.

Consider a RPB as a basic one and let us start unfolding it around one of its arbitrarily chosen vertex completing the
respective local branching structure of the RPRS defined by this vertex. Such a vertex unfolding will be called {\it complete}.
Note that none of the obtain emages of the polygon is faithful with respect to any other of them in any complete vertex unfolding.
Note also that if the angle with the chosen vertex is equal to $\frac{p_k}{q_k}\pi$ then the complete unfolding considered is
invariant under rotations by integer multiples of the angle  $\frac{2p_k}{q_k}\pi$. Such invariant rotations will be called further
{\it vertex rotations} around a given vertex.

Next continue unfoldings around the remaining vertex
of the polygon in a similar way to obtain a compact and connected figure with the following two properties
\begin{itemize}
\item none of the polygon image contained by the figure is faithful with respect to any other of them belonging to the figure; and
\item any other polygon image obtained by the further unfolding process becomes a faithful picture of some of the polygons
belonging to the figure.
\end{itemize}

It is clear that such a figure will contain a finite
number of the basic polygon images obtained by a finite even number of reflections. Each such a figure
will be called {\it elementary polygon pattern} (EPP).

EPP's are however not unique. Nevertheless all they possess the following properties
\begin{itemize}
\item a number $2C$ of polygon emages they contained is the same for all of them and is even;
\item their boundaries are constituted by the sides of emages of the unfolded polygon;
\item sides of an EPP appear in pairs of parallel sides, i.e. a subsequent reflection of the polygon by one of these sides
gives as an image the faithful one of the polygon containing the second side; and
\item  identifying any such a pair of sides of EPP one can conclude that a number of all emages of {\it any} side of the polygon in EPP is the same
and equal to $C$.
\end{itemize}

Consider an EPP and a pair of polygons which sides are pieces of the EPP boundary and are parallel to each other. Such sides can be joined
by a period of the respective RPRS. Such a period will called {\it simple} for a given EPP. Translating by this period any polygon of the
considered polygon pair to cover their parallel sides we get another
EPP. Any EPP can be obtained from the other ones in this way, i.e. by successive simple period translations of polygons with
parallel sides. It is obvious that
\begin{itemize}
\item such simple period translations leave the total number of polygons unchanged in successive EPP's;
\item by respective simple period translations one can reconstruct each complete unfolding of a RPB around any of its vertex; and
consequently
\item since $C$ is the number of any polygon side emages in EPP and $k$-th vertex which is enclosed by an angle $\frac{p_k}{q_k}\pi$ provides us
with a complete unfolding around the vertex containing $q_k$ emages of such a side then $C=n_kq_k$ where $n_k$ is the total number of the $k$-th vertex
emages in EPP, i.e. $C$ has to be the least common multiple of all $q_k$, $k=1,...,n$.
\end{itemize}

Any two EPP's related by the above simple period translations of polygons they contained will be called {\it equivalent}. A number of
all equivalent EPP's is of course finite.

Each EPP is a periodic element of RPRS reconstructing it completely by its all simple periodic translations and their multiple repetitions.

Simple periods of an EPP can be of two kinds
\begin{itemize}
\item the ones which can be represented by translation vectors lying totally {\it inside} a EPP; and
\item the remaining ones.
\end{itemize}

The first kind of simple periods provide us with periods of periodic trajectories, i.e. each such a trajectory is equipped with a period of this kind
and each such a simple period defines a periodic trajectory and bundles of them as well with this simple period.

The second kind of simple periods which will be called {\it structural} act between two {\it different} sheets of RPRS, i.e. a trajectory
starting from one side of EPP and running
in the direction of such a period to the corresponding parallel side cannot cross the latter. It means of course that such a trajectory
passes close to a corresponding branching point generating the sheets. Therefore this kind of periods has a structural nature for RPRS
representing its periodic structure.

Among all possible EPP's one can distinguish those which are invariant under the vertex rotations around some of the RPB vertices. It should
be clear that starting from any such a vertex one can always built a corresponding invariant EPP. Such EPP's will be called {\it rotationally
invariant}. If such a rotationally invariant EPP is generated by a vertex with the corresponding angle equal to $\frac{p_k}{q_k}\pi$ then
the respective invariant rotations are defined by integer multiples of $\frac{2p_k}{q_k}\pi$

Consider an EPP. Identifying each pair of their corresponding parallel sides
we transform the EPP into a two-dimensional closed compact surface of a genus $g$ given by (see for example \cite{8,29,6}):
\be
g=1+\frac{C}{2}\sum_{k=1}^n\frac{p_k-1}{q_k}
\label{1}
\ee
where $n$ is a number of the polygon vertices.

\begin{figure}
\begin{center}
\psfig{figure=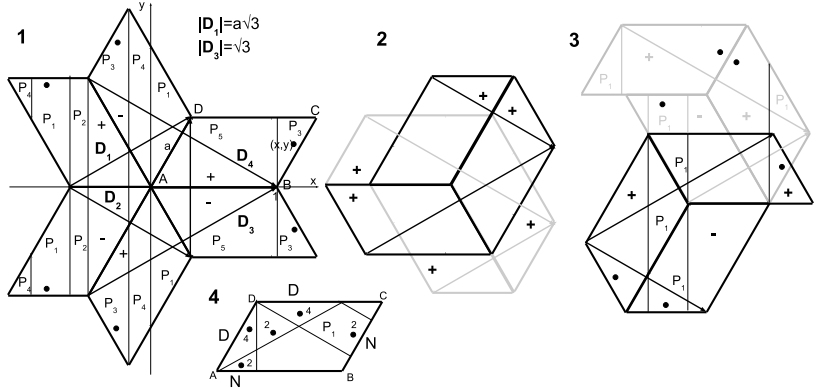,width=11cm} \caption{Different EPP's (Fig.Fig.2.1 - 2.3) for the parallelogram ABCD with the smaller angle equal to $\pi/3$.
The four independent periods same for each EPP are shown as well as five POC's parallel to the period ${\bf D}_1-{\bf D}_2$.
The prescription of signs is consistent for the first two EPP's and inconsistent for the third one (see sec.4.1). A SWF (superscar)
built in
the parallelogram ABCD (Fig.2.4) on the POC $P_1$ according to the sign prescriptions shown in the third EPP (Fig.2.3) satisfies the Dirichlet and
the Neumannn boundary conditions marked by $D$ and $N$ on respective segments of the parallelogram boundary (see sec.6). Numbers at the
distinguished points of Fig.2.4 show the multiplicity by which the $P_1$-POC flow covers the points.}
\end{center}
\end{figure}

Note that the obtained surface is independent of the chosen EPP.

The respective constructions of two EPP's for the parallelogram with the smaller angle equal to $\pi/3$ is shown in
Fig.2

Let us project now all the periods of RPRS on the plane occupied by the original RPB. Then the set of them has the
following two properties:
\begin{enumerate}
\item at most only two of them are linearly independent in the algebra of the real numbers; and
\item at most only $2g$ of them are linearly independent in the algebra of integer numbers.
\end{enumerate}

Let ${\bf D}_k,\;k=1,...,2g$, be periods which are linearly independent in the algebra of integer numbers. Then we
have for any period {\bf D}:
\be
{\bf D}=n_1{\bf D}_1+ ... +n_{2g}{\bf D}_{2g}
\label{2}
\ee

The last relation defines the period algebra on the plane with an infinite but countable number of elements.

A fundamental role of EPP's can be expressed now as the statement that all linearly
independent periods of RPRS are contained in the set of all simple periods of all equivalent EPP's.

Let us however note further that a period which is simple in an EPP can be not as such in another equivalent EPP, i.e. to be reconstructed
in the latter EPP
it needs to be "broken" in some number of pieces to join the corresponding parallel edges of the EPP. Such a "broken" period in an EPP will
be called {\it compound}. It means therefore that compound periods which can found in an EPP can be identified as a simple one in another
EPP equivalent to the former.

In fact having only a {\it single} EPP and identifying {\it all} its independent periods both the simple and the compound ones one can collect
all linearly independent periods of RPRS.

\subsection{Periodic orbit channels of Bogomolny and Schmit}

\hskip+2em Consider now a relation between periods of a RPRS which are not structural and branch points of the latter formed by the branching polygon
vertices with $p_k>1$ in \mref{1}. For this goal consider a bundle of periodic rays with such a period.
Such a bundle has been called {\it a periodic orbit channel} (POC) by Bogomolny and Schmit. The POC cannot pass however by any
branching vertex of RPRS since it would have to be splitted
into two another periodic bundles with two in general different periods, i.e. branching polygon vertices of the RPRS can lie only
on the periodic bundle boundary.

Therefore POC's with different although parallel periods (i.e. not related by \mref{2}) have to lie on different sheets of the RPRS.

\subsection{A real algebra of periods and doubly rational polygon billiards}

\hskip+2em While the $2g$ periods ${\bf D}_k,\;k=1,...,2g$, are linear independent in the algebra of the integer numbers they are not
as such in the real number algebra so that taking a pair ${\bf D}_k,\;k=1,2$ of them being linear independent we can
represent the remaining periods ${\bf D}_k,\;k=3,...,2g$ by the following relations:
\be
{\bf D}_k=a_{k1}{\bf D}_1+a_{k2}{\bf D}_2,\;\;\;\;k=3,...,2g
\label{3}
\ee
where $a_{ki},\;i=1,2,\;k=3,...,2g$, are real numbers some of them or even all can be irrational.

It is convenient to notice that by applying the relations \mref{2} we can always reduce all the periods ${\bf D}_k,\;k=3,...,2g$ to lie in the
parallelogram defined by the periods ${\bf D}_k,\;k=1,2$, so that for the number $a_{ki},\;i=1,2,\;k=3,...,2g$ we can have
\be
0\leq a_{ki}\leq 1\;\;\;\;\;\nn\\
i=1,2,\;k=3,...,2g
\label{3b}
\ee

Consider all linear combinations $n_1{\bf D}_1+n_2{\bf D}_2,\;n_1,n_2=\pm 1,\pm 2,...,$ of the periods
${\bf D}_k,\;k=1,2$. They produce on the plane a regular lattice of points.

However it is not difficult to note that translating this lattice by all integer
multiples of every of the remaining periods covers densely the plane by the vertices of the translated lattice if the numbers
$a_{ki},\;i=1,2,\;k=3,...,2g$ are not {\it rational} all, i.e. a set of points on the plane defined by all the periods \mref{2} will be
dense on the plane in such a case.

The last conclusion means that in such a case there is no room for quantizing the motion in the corresponding RPB
semiclassically by the Maslov - Fedoriuk method \cite{4}.

To discuss this problem further let us note that {\it any} polygon is fixed by the sequence $(\alpha_1,...,\alpha_n,{\bf a}_1,...,{\bf a}_n)$ which
elements are defined by Fig.1 so that the vectors ${\bf a}_k,\;k=1,...n$, forms a {\it closed} chain of them. Consider the vectors to behave
as pseudo-ones by the polygon reflections. If the polygon is rational then an EPP
corresponding to it is also represented by closed chains of the vectors ${\bf a}_k,\;k=1,...n$ and their emages. Formally the inner EPP vectors
annihilate each other on each polygon side except the vectors forming a boundary of the EPP itself. Let us however ignore all such annihilations.
Then any simple period of the EPP can be represented as a {\it vector sum} constructed from properly chosen EPP vectors.

Let us now transform the considered polygon  $(\alpha_1,...,\alpha_n,{\bf a}_1,...,{\bf a}_n)$ into the following
$(\alpha_1,...,\alpha_n,{\bf b}_1,...,{\bf b}_n)$, i.e. keeping the polygon angles {\it fixed}. It is then clear that if the vectors
${\bf a}_1,{\bf b}_1$ are parallel to each other then the remaining vectors of both the polygons
with the same indices are also parallel to each other. Let us call such two polygons {\it angle similar}. It is then clear that EPP's
corresponding to both the polygons are also pairwise angle similar so are the vectors constituted them, i.e. they are also pairwise parallel
to each other.

To see how a freedom we have considering families of angle similar polygons (ASP) let us note that lengths $a_k,\;k=1,...,n$, of the sides of the
general polygon of Fig.1 are related by the following two constraints
\be
\sum_{k=1}^n(-1)^{k-1}a_k\cos(\alpha_1+...+\alpha_{k-1})=0\nn\\
\sum_{k=1}^n(-1)^ka_k\sin(\alpha_1+...+\alpha_{k-1})=0\nn\\
\alpha_0=0
\label{3a}
\ee
i.e. there are $n-2$ independent lengths which can be freely changed to form a full family of ASP's. Of course this freedom
can be limited by other constraints put on the polygons.

However the mentioned earlier parallelness property of the sides of ASP's is {\it not} in general shared by simple periods of both the EPP's which
changes in general both their directions and values by angle similar
transformations. This makes a hope that by such transformations of a RP one can get such its angle similar form for which in the
relations \mref{3} all the coefficients $a_{ki},\;i=1,2,\;k=3,...,2g$ become rational. If it is the case we shall call the respective polygon
{\it doubly rational} (DRP).

It is easy to note that from {\it any} broken rectangle \cite{38,3} one can always obtain infinitely many doubly rational ones by its
angle similar transformations, see Fig.3. The same is true also for some rational parallelograms such as the ones shown in Fig.2 the semiclassical
quantization of which is discussed in sec.5. The broken forms of the latter billiards shown in Fig.4 can be also doubly rational.

Moreover in the cases mentioned above the sets of angle similar DRP's are {\it dens} in the respective sets of all angle similar polygons
having the forms of parallelograms or the forms of both the broken rectangles and the broken parallelograms.

It is therefore reasonable to distinguish among all RPB's the families of doubly rational polygon billiards (DRPB) as sets of all angle
similar RP's which contain dens subsets of DRP's. As it will be shown in next sections the main striking feature of the doubly rational
billiards is that their semiclassical quantization is {\it exact}.

Consider now a DRPB. Then all $a_{ki}$ in \mref{3} are rational, i.e.
\be
a_{ki}=\frac{p_{ki}}{q_{ki}}
\label{4}
\ee
where $p_{ki},q_{ki}$ are coprime integers with $0\leq p_{ki}\leq q_{ki}$.

Let further $C_1$ be the lowest common multiple of $q_{k1}$ and $C_2$ of the respective $q_{k2}$,
i.e. $C_i=q_{ki}n_{ki},\;i=1,2,\;k=3,...,2g$. Then the relations \mref{3} can be rewritten as
\be
{\bf D}_k=n_{k1}p_{k1}\frac{{\bf D}_1}{C_1}+n_{k2}p_{k2}\frac{{\bf D}_2}{C_2},\;\;\;\;k=3,...,2g
\label{5}
\ee
and
\be
{\bf D}=n_1{\bf D}_1+ ... +n_{2g}{\bf D}_{2g}=
(n_1q+n_3n_{31}p_{31}+...+n_{2g}n_{2g1}p_{2g1})\frac{{\bf D}_1}{C_1}+\nn\\
(n_2q+n_3n_{32}p_{32}+...+n_{2g}n_{2g2}p_{2g2})\frac{{\bf D}_2}{C_2}=r_1\frac{{\bf D}_1}{C_1}+r_2\frac{{\bf D}_2}{C_2}
\label{6}
\ee
for any period {\bf D} of RPRS while $r_j,\;j=1,2$, are integers and $C_i,\;i=1,2$, are {\bf D}-independent. Of course
$r_i=C_i$ for ${\bf D}_i,\;i=1,2$.

The relations \mref{6} show therefore that all periods of the DRPB's can be done
mutually commensurate being all linear combinations of two independent vectors
$\frac{{\bf D}_i}{C_i},\;i=1,2$, in the algebra of integer numbers.

Obviously the form of the relation \mref{6} is independent of a choice of the linearly independent pair of periods but
the factors $C_i,\;i=1,2$, can of course depend on such a choice.

Note also that if the classical motion in the RPB is integrable then $C_i=1,\;i=1,2$, since each period {\bf D} satisfies then \mref{1}
with $g=1$.

The above representation of the DRPB periods will be utilized in the semiclassical quantization of motions in DRPB.

\subsection{Unfolding skeletons and semiclassical wave functions defined on DRPB's}

\hskip+2em Consider a rational polygon billiard in its basic position and choose a global skeleton defined on it, Fig.1.
According to our description of the global skeleton it can be defined by some of its $2C$ global compound bundles. Suppose that such
a bundle has been chosen. Then all rays it contains are parallel to each other. Let us unfold the
polygon considered infinitely in every direction together with the rays of the global bundle chosen. It is clear that
all the rays of the bundle will be transformed into an infinite family of straight lines parallel to the rays of the
global bundle but totally covering the RPRS.

Assuming some coordinate system on the RPRS suppose further that we have constructed in the considered RPB on the chosen
global skeleton a semiclassical wave function (SWF)
$\Psi^{as}(x,y,p_x,p_y)$ with the {\it classical} momentum ${\bf p}^{cl}=(p_x,p_y)$ parallel to the rays of the global
bundle chosen. It is
clear that by unfolding the polygon the SWF is extended on the whole RPRS. Because of that it becomes a periodic
function defined on the RPRS with the periods considered in the previous section.

However since on the plane any function
can be periodic with at most two independent periods (in the algebra of integer numbers) then to avoid the obvious
contradiction for the case considered we have to focus ourselves on DRPB's to demand from
$\Psi^{as}(x,y,p_x,p_y)$ to be periodic under {\it any} period of the corresponding RPRS, i.e.
\be
\Psi^{as}(x+D_x,y+D_y,p_x,p_y)=\Psi^{as}(x,y,p_x,p_y)
\label{7}
\ee
for any period ${\bf D}=(D_x,D_y)$ of the RPRS.

According to its construction \cite{38,3} a SWF is a sum of the basic semiclassical wave functions (BSWF) of the form
\be
\Psi^\pm(x,y,p_x,p_y)=e^{\pm i\lambda(p_xx+p_yy)}\chi^\pm(x,y,p)
\label{8}
\ee
where $\lambda=\hbar^{-1}$ (and will be
put further equal to 1 as well as the billiard ball mass), $p$ is a value of the billiard ball classical momentum
 ${\bf p}^{cl}$ and the factors $\chi^\pm(x,y,p)$ are given by the following semiclassical series
for $p\to +\infty$:
\be
\chi^\pm(x,y,p)=\sum_{k\geq 0}\frac{\chi_k^\pm(x,y)}{p^k}
\label{9}
\ee

It is clear that the BWSF's have to be also periodic on the RPRS satisfying the equations
\be
e^{\pm i(p_x(x+D_x)+p_y(y+D_y))}=e^{\pm i(p_xx+p_yy)}e^{\pm i{\bf p}^{cl}\cdot{\bf D}}=e^{\pm i(p_xx+p_yy)}\nn\\
\chi_k^\pm(x+D_x,y+D_y)=\chi_k^\pm(x,y),\;\;\;\;k\geq 0
\label{10}
\ee
for any period ${\bf D}=(D_x,D_y)$ of the RPRS so for the classical momenta we have to
have the conditions
\be
{\bf p}^{cl}\cdot{\bf D}=2k\pi,\;\;\;\;\;\;\;\;k=0,\pm1,\pm2,...
\label{11}
\ee

To solve the conditions \mref{10} and \mref{11} we have to consider further two different cases of the global skeletons, namely
the ones which trajectories are parallel to some of the periods of the corresponding RPRS and contain periodic trajectories
and the remaining ones. The
first kind of the global skeletons, which will be called periodic, have to contain of course POC's while the second kind, the
aperiodic ones, are completely deprived of any periodic trajectories.

\subsubsection{Periodicity constraints put on a momentum of a motion on aperiodic skeletons}

\hskip+2em Consider first an aperiodic global skeleton. It means that a momentum ${\bf p}^{cl}$ of the billiard ball moving on the
skeleton cannot be parallel to any of the periods of RPRS so that it has two independent projections on the periods
${\bf D}_i,\;i=1,2$.

Let us now make use of the commensurateness of the periods expressed by \mref{6} and enforce the BSWF's \mref{8}
to be periodic with two periods equal to $\frac{{\bf D}_1}{C_1}$ and $\frac{{\bf D}_2}{C_2}$. It is clear that then these
BSWF's will be periodic with respect to all periods of the RPRS. Therefore we demand for the classical
momenta to satisfy the following conditions
\be
{\bf p}^{cl}\cdot{\bf D}_1=2\pi mC_1,\;\;\;\;\;\;\;\;\;\;\;\;\;\;\;\;\;\;\;\;\;\;\;\;\;\;\;\;\;\;\;\;\;\;\;\;\;\nn\\
{\bf p}^{cl}\cdot{\bf D}_2=2\pi nC_2,\;\;\;\;\;\;\;m,n=0,\pm 1,\pm 2,...
\label{12}
\ee
and hence
\be
{\bf p}_{mn}^{cl}=2\pi\frac{(mC_1{\bf D}_2-nC_2{\bf D}_1)\times({\bf D}_1\times{\bf D}_2)}{({\bf D}_1\times{\bf D}_2)^2}=\nn\\
2\pi\frac{mC_1D_2^2-nC_2{\bf D}_1\cdot{\bf D}_2}{({\bf D}_1\times{\bf D}_2)^2}{\bf D}_1+
2\pi\frac{nC_2D_1^2-mC_1{\bf D}_1\cdot{\bf D}_2}{({\bf D}_1\times{\bf D}_2)^2}{\bf D}_2\nn\\
|m|+|n|>0,\;\;\;\;\;\;\;\;\;\;\;\;m,n=0,\pm 1,\pm 2,...
\label{13}
\ee
i.e. the possible classical momenta of the billiard ball have to be quantized just by the periodic structure
of the RPRS only.

It is to be noted that both the form of the formula \mref{13} and the momentum spectra it provides are independent of
the choice of the linear independent pair
of periods $D_i,\;i=1,...,2g$. However a knowledge of any pair of these periods is not sufficient for the formula
\mref{13} to be completed, i.e. for that goal the formula needs the constants $C_i,\;i=1,2,$ to be known also and the
latter can be established only when the remaining independent periods are also identified.

Let $p_i,\;p$ be projections of the momentum ${\bf p}^{cl}$ on the periods ${\bf D}_i,\;i=1,2$, and ${\bf D}$ as given by \mref{6} respectively.
Then the conditions \mref{12} and the relation \mref{6}
can be written in terms of the respective wave lengths $\lambda_i=\frac{2\pi}{|p_i|},\;i=1,2$ and $\lambda=\frac{2\pi}{|p|}$ as follows
\be
\lambda_i=\frac{D_i}{n_iC_i}\nn\\
\lambda=\frac{D}{r_1n_1+r_2n_2}\nn\\
\;\;\;\;\;\;\;n_i=1, 2,...,\;i=1,2
\label{12a}
\ee
where $D,\;D_i$ are lengths of the periods ${\bf D},\;{\bf D}_i,\;i=1,2$.

It follows from \mref{12a} that each period of RPRS measured by the respective wave length has the integer total length.

\subsubsection{Periodicity constraints put on the classical momentum of a motion on a global periodic skeleton}

\hskip+2em Consider a global periodic skeleton which contains a periodic trajectory with the period ${\bf D}_2$. Then
corresponding classical momentum is also parallel to this period so that taking into account \mref{13} we have the following
quantization condition for the classical momentum of the periodic skeleton
\be
mC_1D_2^2-nC_2{\bf D}_2\cdot{\bf D}_1=0,\;\;\;\;\;\;\;m=0,\pm 1,\pm 2,...,\;n=\pm 1,\pm 2,...
\label{12b}
\ee
and
\be
{\bf p}_n^{per}=\frac{2\pi nC_2}{D_2^2}{\bf D}_2,\;\;\;\;\;\;\;n=\pm 1,\pm 2,...
\label{12c}
\ee
The condition \mref{12b} is of course a constraint on the periods ${\bf D}_1$ and ${\bf D}_2$ with the following
solution independent of $m$ and $n$
\be
C_2{\bf D}_2\cdot{\bf D}_1=kC_1D_2^2\;\;\;\;\;\;\;\;\;\;\;\;\;\;\;\;\;\;\;\;\;\;\;\;\;\;\;\;\nn\\
m=kn,\;\;\;\;\;\;\;\;\;\;\;\;\;\;\;\;\;\;\;\;\;\;\;\;\;\;\;\;\;\;\;\;\nn\\
k=0,\pm 1,\pm 2,...,\;n=\pm 1,\pm 2,...
\label{12d}
\ee

The last condition however cannot be satisfied in general
by an arbitrary RPB which defines the periods ${\bf D}_i,\;i=1,2$, uniquely. However for some particular RPB's and for some particular $k$
\mref{12d} can be satisfied. Such a possibility takes place for example if the periods ${\bf D}_i,\;i=1,2$, are perpendicular to each other so
that $k=0$ then. Nevertheless the conditions \mref{12d} limit possible forms of RPB seriously.

RPB's for which there are no any pair of linear independent periods satisfying the conditions \mref{12d} for some integer $k$ will
be called {\it generic}.

We can conclude therefore that for generic cases of RPB's only aperiodic skeletons provide us with a
possibility of consistent construction on them of semiclassical eigenfunctions of energy together with the corresponding semiclassical
spectra of the latter. For the non-generic forms of RPB's it is necessary to consider also global periodic skeletons to quantize the
corresponding classical motions fully.

The global periodic skeleton considered contains of course at least one POC with the period ${\bf D}_2$. If there are more POC's with
the periods ${\bf D}_l,\;l=3,...,r$, then ${\bf D}_l=p_l/q_l{\bf D}_2$ and $C_2=q_ln_l$ for integer $n_l,\;l=3,...,r$, so that
\be
{\bf p}_n^{per}\cdot{\bf D}_l=2\pi nC_2\frac{p_l}{q_l}=2\pi nn_lp_l
\label{12e}
\ee
and if $C_l\equiv n_lp_l$ then
\be
{\bf p}_n^{per}=\frac{2\pi nC_l}{D_l^2}{\bf D}_l,\;\;\;\;\;\;\;n=\pm 1,\pm 2,...,\;l=3,...,r
\label{12f}
\ee

\section{Energy quantization on skeletons in DRPB's}

\hskip+2em  Let us choose the $x,y$-coordinates on the RPRS to be such that the $y$-axis is parallel to the rays of the
considered unfolded skeleton so that the $x$-axis is
perpendicular to the rays. Any such a coordinate system will be called {\it local} for the considered skeleton.

The factors
$\chi^\sigma(x,y,p),\;\sigma=\pm$, of the BSWF's \mref{8} have then to satisfy the following semiclassical limit
$p\to+\infty$ of the Schrödinger equation \cite{38,3}
\be
\sigma 2ip\frac{\p\chi^\sigma(x,y,p)}{\p y}+\triangle \chi^\sigma(x,y,p)+
2(E-\fr p^2)\chi^\sigma(x,y,p)=0
\label{14}
\ee
where $E$ is the energy parameter.

Note also that the variable $x$ enumerates locally the rays of the skeleton.

In the semiclassical limit $p\to+\infty$ we are looking for the semiclassical spectrum of the billiard ball energy $E$
in the form of the following semiclassical series \cite{38,3}
\be
E=\fr p^2+\sum_{i\geq 0}\frac{E_k}{p^k}
\label{15}
\ee

Using \mref{9} and \mref{15} the equation \mref{14} can be solved recurrently to get \cite{38,3}
\be
\chi_0^\sigma(x,y)\equiv\chi_0^\sigma(x)\nn\\
\chi_{k+1}^\sigma(x,y)=\chi_{k+1}^\sigma(x)+
\frac{\sigma i}{2}\int_0^y\ll(\triangle\chi_{k}^\sigma(x,z)+
2\sum_{l=0}^kE_{k-l}\chi_{l}^\sigma(x,z)\r)dz\nn\\
k=0,1,2,...
\label{16}
\ee

\subsection{Energy quantization on aperiodic global skeletons}

\hskip+2em Consider a generic RPB and a particular momentum quantized in it according to \mref{12}, i.e. corresponding to an
aperiodic global skeleton which starts from some basic polygon. All its trajectories start from a definite part of the
polygon boundary to move by the RPRS. Let us choose any of its trajectory and follow its running on the RPRS. By its
aperiodicity the trajectory meeting faithful images of the basic polygon never cuts its boundary in the same point from
which it starts. In fact since every trajectory meets on its way infinitely many faithful images it cuts their boundary
in points which if collected together are densely distributed on the starting boundary of the basic polygon.

According to the formula \mref{16} the factors $\chi^\sigma(x,y,p),\;\sigma=\pm$, of the BSWF's \mref{8} for a given
trajectory change only along it just by varying $y$. However its zeroth order term $\chi_0^\sigma(x)$ does not depend on
$y$.
Therefore its value on a given trajectory is distributed densely on others and demanding its continuity on the polygon
boundary we come to the conclusion that it has to be a constant function of $x$ on the polygon boundary.

From the
recurrent relations we get immediately that the same property have to have the remaining terms of the semiclassical
series \mref{9}, i.e. the factors $\chi^\sigma(x,y,p),\;\sigma=\pm$, have constant values independent of $x$ and $y$.
Moreover the corresponding energy coefficients $E_k,\;k\geq 0$, of the semiclassical expansion for energy \mref{15} have
all to be equal to zero in such a case.

Therefore we can put both the factors $\chi^\sigma(x,y,p),\;\sigma=\pm$, equal to unity and to conclude that in the generic
cases of RPB's the BSWF's defined on them have to be constructed in \mref{8} by the exponential factors only and the
energy spectrum is then given by
\be
E_{mn}=\frac{\ll({\bf p}_{mn}^{cl}\r)^2}{2}=2\pi^2\frac{|mC_1{\bf D}_2-nC_2{\bf D}_1|^2}{|{\bf D}_1\times{\bf D}_2|^2}
,\;\;\;\;\;\;\;m,n=\pm 1,\pm 2,...
\label{17}
\ee

\subsection{Energy quantization on global periodic skeletons}

\hskip+2em If a RPB is not generic then there are at least two linear independent periods say ${\bf D}_i,\;i=1,2$ which
for some integer $k$ satisfy the relation \mref{12d}. The quantization condition \mref{13} takes then the form
\be
{\bf p}_{mn}^{cl}=
2\pi\frac{nC_2D_1^2-mC_1{\bf D}_2\cdot{\bf D}_1}{({\bf D}_1\times{\bf D}_2)^2}{\bf D}_2+
2\pi\frac{(m-kn)C_1D_2^2}{({\bf D}_1\times{\bf D}_2)^2}{\bf D}_1\nn\\
|m|+|n|>0,\;\;\;\;\;\;\;\;\;\;\;\;m,n=0,\pm 1,\pm 2,...
\label{18}
\ee
which reduces to \mref{12c} for $m=kn$, i.e. ${\bf p}_n^{per}\equiv{\bf p}_{kn\;n}^{cl}$.

In the considered case there are of course global aperiodic skeletons but also global periodic ones one of which
has a momentum parallel to the period ${\bf D}_2$ which can be quantized according
to the conditions \mref{12c} and \mref{12d}.

For global aperiodic skeletons the energy spectrum is still given by \mref{17} where momenta ${\bf p}_{mn}^{cl}$ are given
by \mref{18}, i.e.
\be
E_{mn}^{ap}=\frac{\ll({\bf p}_{mn}^{cl}\r)^2}{2}=2\pi^2\frac{m^2C_1^2D_2^2+n^2C_2^2D_1^2-2mnkC_1^2D_2^2}
{|{\bf D}_1\times{\bf D}_2|^2},\;\;\;\;\;\;\;m,n=\pm 1,\pm 2,...
\label{18a}
\ee

For a global periodic skeleton however the corresponding energy spectrum is a sum of the classical kinetic energy given by \mref{17}
and the remaining terms of the semiclassical series \mref{15}. The latter have to be established by solving the equations
\mref{15} and \mref{16} for the case of the periodic skeleton considered. This has been done in app.B.

As it follows from app.B BSWF's in POC's and in aperiodic bundles of the global periodic skeleton differ in their forms. These
differences are essential for global BSWF's which have to be constructed by a smooth matching of the BSWF's defined on POC's and on aperiodic
bundles of the skeleton.

Making such a matching of BSWF's between any two neighbor POC's we conclude that
on their common boundaries $p_n^{per}$ and $E_{k,0}$ have to be the same for each POC.

Matching however two BSWF's on a common boundary of a POC and an aperiodic bundle we are led to the conclusion that $E_{k,0}=0$ for each POC,
and $p_n^{per}$ has the same value for all the bundles of the skeleton. i.e. the considered periodic skeleton behaves in such a case as an
aperiodic one.

As it follows from app.B corresponding forms of the BSWF's in the global periodic skeleton written in its local
$x,y$-variables can be therefore the following
\be
\Psi_{mn}^\pm(x,y)=e^{\pm i(\pm\sqrt{2E_{0,m}}x+p_n^{per}y)}\nn\\
\sqrt{2E_{0,m}}D_1\sin\alpha=2mC_1\pi,\;\;\;\;\;\;\;\;m=0,1,2,...\nn\\
\label{19}
\ee
where the case $m=0$ corresponds to the presence at least one of aperiodic bundles in the skeleton while the remaining values of $m$
correspond to their total absence in the skeleton, i.e. the skeleton is constructed then only of POC's. In the above
formula $\alpha$ is the angle between the periods ${\bf D}_1$ and ${\bf D}_2$, $E_{0,m}$ and $p_n^{per}$ are the same for all
POC's and aperiodic bundles and $\pm$-signs in \mref{19} are independent.

The independence of the form \mref{19} of $\Psi_{mn}^\pm(x,y)$ on POC's is due to linear rational relations between all their
periods as well as due to similar relations between the period ${\bf D}_1$ and the remaining
periods of the RPRS not parallel to the period ${\bf D}_2$ since for them we have
\be
D_x=\frac{p}{q}D_{1,x}\nn\\
\sqrt{2E_{0,m}}D_x=\frac{p}{q}\sqrt{2E_{0,m}}D_1\sin\alpha=2mC_1\pi\frac{p}{q}=2mpr\pi,\;\;\;\;\;\;\;\;m=1,2,...
\label{20}
\ee
since $C_1=rq$ for some integer $r$.

Therefore for the energy spectrum generated by the global periodic skeleton defined by the periods ${\bf D}_1$ and
${\bf D}_2$ we get
\be
E_{mn}^{per}=\fr \ll(p_n^{per}\r)^2+E_{0,m}=2\pi^2\ll(\frac{m^2C_1^2}{D_1^2\sin^2\alpha}+\frac{n^2C_2^2}{D_2^2}\r)
=\nn\\
2\pi^2\frac{m^2C_1^2D_2^2+n^2C_2^2D_1^2-k^2n^2C_1^2D_2^2}{|{\bf D}_1\times{\bf D}_2|^2}\nn\\
m,n=1,2,...
\label{21}
\ee

It is to be noted that despite an apparent similarity between the energy spectrum formulae for the global periodic
skeletons \mref{21} and the aperiodic ones \mref{18a} the formulae are in general different. They coincide only for
$k=0$, i.e. when the periods  ${\bf D}_1$ and ${\bf D}_2$ are orthogonal to each other.

However there are also other essential differences between both the cases.

The first one follows from the fact that
in the spectra \mref{21} $E_{0,m}$ is the second term of the semiclassical expansion for the energy and therefore it
should be clearly smaller than the first one, i.e. it has to satisfy the following inequality
\be
\sqrt{2E_{0,m}}<<p_n^{per}
\label{21b}
\ee
or
\be
|m|<<\frac{C_2D_1\sin\alpha}{C_1D_2}|n|\nn\\
m,n=1,2,...
\label{21c}
\ee

The last condition is just the one which has justified the considerations of Bogomolny and Schmit on the superscar phenomenon \cite{1,2}.

There is no a relation like \mref{21c} for the aperiodic case spectra for which the unique condition is that $p_{mn}^{cl}$ has to be large
(in comparison with $p_{11}^{cl}$, for example) the latter condition being also satisfied by the periodic case spectra.

The second difference between the aperiodic and the periodic cases is that in the latter case all the SWF's can be
built on the {\it same} global periodic skeleton independently of the momenta (which are {\it always} parallel to the
period ${\bf D}_1$) while in the opposite case for different momenta ${\bf p}_{mn}^{cl}$ the corresponding aperiodic skeletons are
{\it different}.

It will be convenient for further considerations to unify the momentum $p_n^{per}$ and the quantities $\pm\sqrt{2E_{0,m}}$ in the
formula \mref{19} for the global BSWF's
into a pseudo-momentum ${\bf p}_{mn}^q$ having in the local coordinate system of the skeleton the components
$(\pm\sqrt{2E_{0,m}},p_n^{per})$ and which will be called a quantum momentum. By this unification \mref{21} takes the
form similar to the aperiodic cases \mref{17} and \mref{18a}, i.e.
\be
E_{mn}^{per}=\fr \ll({\bf p}_{mn}^q\r)^2
\label{21a}
\ee
under the conditions \mref{21b}-\mref{21c}.

\section{Semiclassical wave function constructed on skeletons\\ in DRPB's}

\hskip+2em We have now to construct SWF's corresponding to the semiclassical energy spectra \mref{17} and \mref{21}. It should be
stressed that these spectra have followed uniquely as the direct consequences of the periodic structure of RPRS's and the
asymptotic structure of BSWF's defined by \mref{8}, \mref{9} and \mref{14}-\mref{16}. Since these
spectra are already fixed they seem to correspond to some particular boundary conditions - the Dirichlet ones, the Neumannn ones
or their mixtures. It is quite surprising that as it will be shown below for the energy spectra mentioned one can easily construct SWF's satisfying
the Dirichlet boundary conditions as well as the Neumannn ones while mixtures of these conditions can be used with some limitations being
even excluded depending on billiard forms.

This last fact i.e. a lack of freedom in choosing mixtures of the Dirichlet and the Neumannn boundary conditions is quite important
since it means that many exact states may have no their representations in the semiclassical
limit relied on the assumption that the classical motions in the RPB's are ruled by the optical reflections of the billiard ball
off the billiard boundary. As an example of such states can be mentioned the symmetric ones in the quantized rhombus billiard which
existence is equivalent to satisfy the Neumannn boundary condition on a single side of the quantized equilateral triangle by the corresponding
SWF's \cite{8}. As a consequence of this is a possible quantum mechanical
incompleteness of the asymptotic states generated by the assumptions on the classical motions in RPB's utilized in our paper.

\subsection{SWF's satisfying desired boundary conditions}

\hskip+2em To construct SWF's satisfying desired boundary conditions we can make use of the EPP's corresponding to a given DRPB
considered as basic.
To do it let us choose an EPP corresponding to this BRPB. Let us enumerate further all $2C$ component polygons of
the EPP prescribing the number one to the BRPB itself. Choose a point inside the BRPB with the coordinates
$(x,y)\equiv (x_1,y_1)$ in the chosen coordinate system. Let further $(x_k,y_k),\;k=2,...,2C$ denote coordinates of all
emages
of the point $(x_1,y_1)$ in the remaining enumerated emages of the BRPB. With every of the polygon of EPP and with the
respective points they contain we can associate any of the signs $\pm$.

Consider further an edge of the considered polygon and all its copies in the chosen EPP including the edge itself. The copies can lie
inside the EPP or can be pieces of the EPP boundary. The latter copies appear always in parallel pairs being translated in each such a pair by
some of the RPRS periods which allows to identify them in each of the pair. Making this we can find that there are exactly $C$
copies of each edge in every EPP.

Let us now note that with each copy of an edge (including the edge itself) is associated a pair of image points, i.e. just the ones which are
reflected in it. It is now important to note that all image points in EPP (its number is $2C$) can be joined in such pairs associated with all
copies of a single edge. Of course component points of such pairs depend on an edge.

Let us now prescribed a definite sign plus or minus to every image point. In this way pairs of points associated with copies of an edge prescribe to
each of them a pair of signs. We say that such a prescription is consistent with respect to this edge if in {\it all} these pairs both signs
are the same, i.e. both are pluses or both are minuses or if in {\it all} these pairs both the signs are opposite. If such a prescription of signs
is consistent with respect to {\it all} edges of the EPP we say that such
a prescription is consistent with respect to the EPP considered.

Note however that if a prescription is consistent for some EPP it is also as such
for all other equivalent EPP's.

Since the image points have been enumerated then we can
associate with the $k^{th}$-image its corresponding sign $\eta_k$ in each consistent prescription. Of course a sign associated with an image
point depends on the prescription used.

The following two prescriptions are consistent with respect to any EPP of any RPB
\begin{enumerate}
\item each pair of signs are opposite for each edge; and
\item all pairs of signs are strictly the same for each edge, i.e. (+,+), by a convention.
\end{enumerate}

The first prescription will be called the Dirichlet one, while the second - the Neumannn one.

Let us note that prescribing the sign "+" to the original point in the BRPB an image point in the Dirichlet prescription gets the sign "+"
if it is obtained by an even number of reflections and the sign "$-$" in the opposite case.

Consider now a global skeleton in the chosen EPP represented by some of its global bundles. Note further that the BSWF's defined in the chosen
global bundle have the same exponential forms $e^{\pm i(Ax+iBy)}$
independently of whether they are defined in the periodic skeletons or in the aperiodic ones where $(A,B)$ are the
components in the chosen coordinate system of the quantum momentum ${\bf p}_{mn}^{q}$ or the quantized classical
momentum ${\bf p}_{mn}^{cl}$
respectively. It is therefore enough to construct with these forms the SWF's satisfying desired boundary
conditions on all the sides of the DRPB unfolded to its EPP which lie inside the EPP while on the sides of the unfolded DRPB
which form the boundary of its EPP the chosen boundary conditions will be satisfied by the periodicity conditions.

For a given EPP consider now a consistent prescription of pairs of signs prescribing the signs $\eta_k,\;k=1,...,2C$, to the image points
of a point $(x,y)=(x_1,y_1)$ of the BRPB so that $\eta_1\equiv+$, by a convention. Then two SWF's with definite boundary conditions on the BRPB edges
are the following
\be
\Psi^{as,\pm}(x,y,A,B)=\sum_{k=1}^{2C}\eta_ke^{\pm i(Ax_k+By_k)}
\label{22}
\ee

The above SWF's have the following properties:
\begin{enumerate}
\item they are uniquely defined in the chosen BRPB;
\item they are smooth inside the BRPB;
\item they satisfy the Dirichlet boundary conditions on these sides (edges) of the BRPB boundary for which the signs of the prescribed pairs
are opposite;
\item they satisfy the Neumannn boundary conditions on these sides (edges) of the BRPB boundary for which the signs of the prescribed pairs
are the same;
\item they are {\it exact}, i.e. they satisfy the Schrödinger equation with the energy spectra \mref{8}, \mref{9} and \mref{14}-\mref{16};
\item they are mutually complex conjugate; and
\item they are independent of the chosen EPP.
\end{enumerate}

In particular for the Dirichlet prescription the corresponding SWF's satisfy the Dirichlet boundary conditions on the BRPB
boundary while for the Neumannn prescription - the Neumannn ones.

Note that if $\Psi^{as,\pm}(x,y,A,B)$ do not coincide with each other (up to a constant) then
the corresponding energy levels $E_{mn}$ are degenerate.

One can rewrite the representation \mref{22} for $\Psi^{as,\pm}(x,y,A,B)$ using the fact that the coordinates $(x_k,y_k),\;k=2,...,2C$, of the
emages of the point $(x,y)$ are linearly dependent on $x$ and $y$ being a result of some rotation of the point $(x,y)$ followed by a translation,
i.e. we can write
\be
x_k=a_{k,x}x+a_{k,y}y+a_k\nn\\
y_k=b_{k,x}x+b_{k,y}y+b_k\nn\\
k=2,...,2C
\label{22a}
\ee

Therefore \mref{22} can take the following form
\be
\Psi^{as,\pm}(x,y,A,B)=\sum_{k=1}^{2C}\eta_ke^{\pm i\alpha_k}e^{\pm i(p_{k,x}x+p_{k,y}y)}\nn\\
p_{1,x}=A,\;p_{1,y}=B\nn\\
p_{k,x}=a_{k,x}A+b_{k,x}A\nn\\
p_{k,y}=a_{k,y}A+b_{k,y}B\nn\\
\alpha_1=0\nn\\
\alpha_k=Aa_k+Bb_k\nn\\
k=2,...,2C
\label{22b}
\ee
where ${\bf p}_k=(p_{k,x},p_{k,y}),\;k=2,...,2C$, are all possible quantized momenta of the billiard ball generated by the quantized
momentum ${\bf p}_1$.

If the considered skeleton is generic then ${\bf p}_1$ coincides with the quantized classical momentum
${\bf p}_{mn}^{cl}$ of the chosen global bundle and the phases $\alpha_k$ are gained by the respective BSWF's along the rays of this global
bundle which after subsequent
reflections off the billiard boundary achieve the point $(x,y)$ with the quantized momentum ${\bf p}_k,\;k=2,...,2C$.

In the
opposite case, i.e. for global periodic skeletons the phases $\alpha_k$ do not seem to have such clear physical interpretation.

\section{Some simple examples of DRPB's - the $\pi/3$-parallelogram billiard and the single bay broken rectangle and broken parallelogram billiards}

\hskip+2em Before considering as an illustration of the DRPB's
let us discuss shortly the simplest cases of the rectangle and equilateral triangle billiards. Both the cases are integrable classically. Our
main interest is in possible boundary conditions which can be satisfied in these billiards. By analyzing the consistent prescriptions of
signs in the corresponding EPP's one finds that in the rectangle billiards despite the Dirichlet and the Neumannn ones there is still possible
to put mixed conditions, i.e. different for different pairs of parallel sides. This exhausts however the allowed possibilities.

In the equilateral triangle however no other possibilities of the sign prescription except the Dirichlet and the Neumannn ones are allowed. This fact
causes the non existence of the symmetric semiclassical states in the rhombus billiard \cite{8} built by using the approach developed
in our paper. This conclusion will be confirmed also by the case of the parallelogram billiard which we are going to consider below.

Let us now come back to the cases of DRPB's mentioned, i.e. to the $\pi/3$-parallelograms shown in Fig.2 and to the single bay broken rectangles
shown in Fig.3.

\begin{figure}
\begin{center}
\psfig{figure=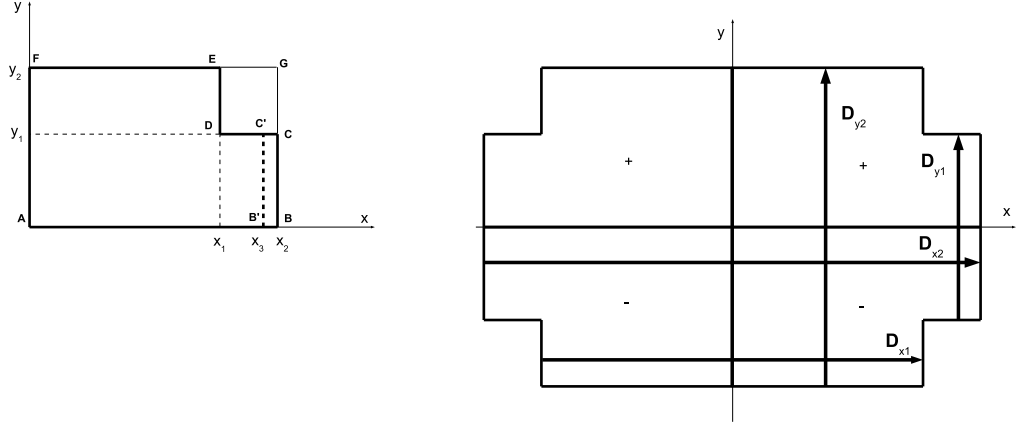,width=14cm} \caption{A single bay broken rectangle billiard $ABCDEF$ (left) with its symmetric EPP (right) on which the four
independent periods of RPRS are shown. An angle similar broken rectangle billiard $AB'C'DEF$ is an $\epsilon$-approximation of the former with
$\epsilon=x_2-x_3$ for $x_2-x_1>1$ or $\epsilon=\frac{x_2-x_3}{x_3-x_1}$ in the opposite case, see app.C}
\end{center}
\end{figure}

\begin{figure}
\begin{center}
\psfig{figure=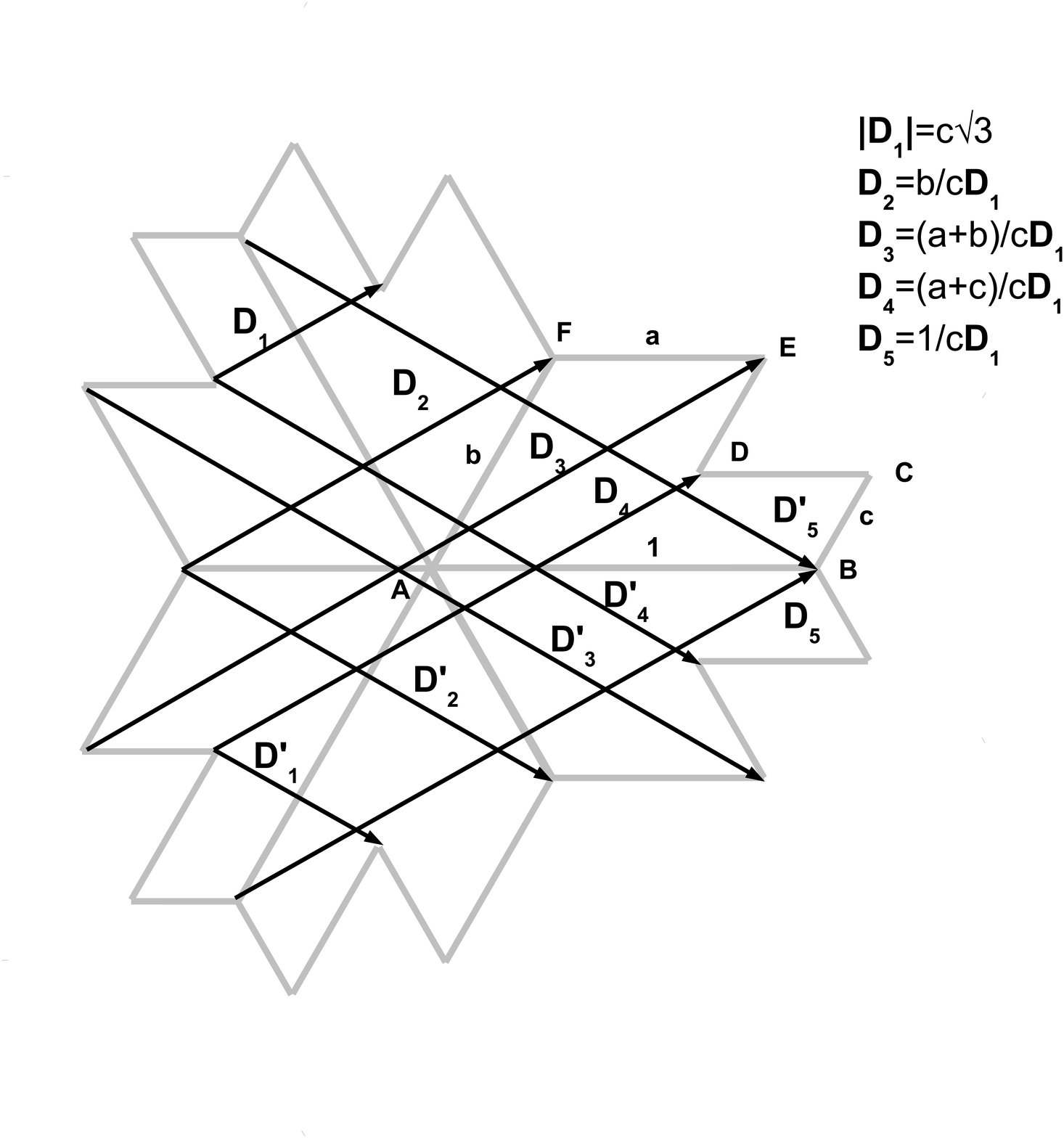,width=8cm} \caption{A symmetric EPP of a single bay broken parallelogram billiard $ABCDEF$ with the smaller angle equal to
$\pi/3$ providing us
with an angle similar family of such billiards containing a dense set od DRPB. There are ten independent periods of the billiards shown in the figure
and corresponding to a multiple torus with $g=5$}
\end{center}
\end{figure}

\subsection{The doubly rational $\pi/3$-parallelogram billiards}

\hskip+2em Considering first the case of the parallelogram one has to note the
four independent periods ${\bf D}_k,\;k=1,...,4$, which are shown in Fig.2 and which are related as follows
\be
{\bf D}_3=\frac{1}{a}{\bf D}_1\nn\\
{\bf D}_4=\frac{1}{a}{\bf D}_2
\label{23}
\ee

It is seen therefore that the angle similar parallelogram billiards considered are doubly rational if $a$ is rational and the set of all these
doubly rational billiards is dense in the set of all the angle similar $\pi/3$-parallelogram billiards. Therefore for doubly rational parallelograms
we have $a=\frac{q}{p}$ so that $C_1=C_2=q$ in the corresponding formulae \mref{5}-\mref{6}.

For the classical momenta and the energy quantized on any generic skeleton in the considered billiard we get then
\be
{\bf p}_{mn}^{cl}=\frac{4\pi p^2}{9q}[(2m-n){\bf D}_1+(2n-m){\bf D}_2]\nn\\
E_{mn}^{gen}=\frac{16}{9}\pi^2p^2(m^2+n^2-mn)\nn\\
m,n=\pm 1,\pm 2,...
\label{24}
\ee

Considering the quantization on periodic skeletons we have to note that there is a pair of periods ${\bf D}_1-{\bf D}_2$ and
${\bf D}_1+{\bf D}_2$ and another pair ${\bf D}_3-{\bf D}_4$ and ${\bf D}_3+{\bf D}_4$ of them in which the respective periods
are perpendicular to each other so that taking them as the new four independent pair of the periods we can
satisfied the corresponding formula \mref{12d} for $k=0$. Then for aperiodic skeletons according to \mref{18} we get for their
quantized classical momenta and energy
\be
{\bf p}_{mn}^{cl}=\frac{2\pi p^2}{9q}[(3m+n){\bf D}_1+(n-3m){\bf D}_2]\nn\\
E_{mn}^{gen}=\frac{4}{9}\pi^2p^2(3m^2+n^2)\nn\\
m,n=\pm 1,\pm 2,...
\label{25}
\ee

Considering further the global periodic skeleton shown in Fig.2 parallel to the period
${\bf D}_1-{\bf D}_2$ and composed of five POC's we get for its quantized classical momentum
\be
{\bf p}_n^{per}=\frac{2\pi np^2}{3q}({\bf D}_1-{\bf D}_2),\;\;\;\;\;\;\;n=\pm 1,\pm 2,...
\label{26}
\ee
and for its energy spectrum
\be
E_{mn}^{per}=\frac{4}{9}\pi^2p^2(3m^2+n^2)\nn\\
|m|<<\sqrt{3}|n|\nn\\
m,n=\pm 1,\pm 2,...
\label{27}
\ee

The formulae \mref{25} and \mref{27} for the energy spectra coincide in accordance with the general rule since $k=0$ for the case considered.
It can be also checked that by the substitutions $m\to m-n$ and $n\to m+n$ these formulae coincide with \mref{24} too because of the same
reason.

We can now construct the SWF's corresponding to the established spectra and to different consistent prescriptions of signs to EPP's.
In Fig.2 are shown two such
consistent prescriptions. It is easy to convince oneself that there are no other ones, i.e. only the Dirichlet prescription and the Neumannn one
are consistent for the parallelogram billiards.

According to the general rule given by \mref{22} we have to find for an original point
of the parallelogram having coordinates $(x,y)$ all coordinates of its image points lying inside the EPP. It is easy to do it using the
first EPP of Fig.2. We collect the corresponding coordinates in pairs symmetric with respect to the $x$-axis. They are
$[(x,y),\;(x,-y)]$, $[(-\fr x+\fr\sqrt{3}y,\fr\sqrt{3}x+\fr y)\;(-\fr x+\fr\sqrt{3}y,-\fr\sqrt{3} x-\fr y)]$ and
$[(-\fr x-\fr\sqrt{3}y,\fr\sqrt{3}x-\fr y),\;(-\fr x-\fr\sqrt{3}y,-\fr\sqrt{3} x+\fr y)]$.

Therefore using the periods ${\bf D}_1-{\bf D}_2$ and ${\bf D}_1+{\bf D}_2$ for the quantization and taking into account that the momenta represented
in the $x,y$-coordinates of Fig.2 by $[A,B]$ are then quantized by
\be
3aA=2\pi(m+n)q\;\;\;\;\;\;\;\;\;\;\;\;\;\;\;\;\;\;\;\;\;\;\;\;\;\;\;\;\;\nn\\
\sqrt{3}aB=2\pi(m-n)q\;\;\;\;\;\;\;\;\;\;\;\;\;\;\;\;\;\;\;\;\;\;\;\;\;\;\;\;\;\nn\\
A^2+B^2\neq 0,\;\;\;\;\;\;m,n=0,\pm 1,\pm 2,...
\label{27a}
\ee
the SWF's satisfying the Dirichlet boundary conditions in the parallelogram billiard are
\be
\Psi^{as,\pm}(x,y,A,B)=e^{\pm iAx}\sin(By)-e^{\pm iA(-\fr x+\fr\sqrt{3}y)}\sin\ll[B(\fr\sqrt{3} x+\fr y)\r]+\nn\\
e^{\pm iA(-\fr x-\fr\sqrt{3}y)}\sin\ll[B(\fr\sqrt{3}x-\fr y)\r]\nn\\
B\neq 0
\label{28}
\ee

One can obtain two real SWF's taking properly the two linear combinations of the above ones, i.e.
\be
\Psi_1^{as}(x,y,A,B)=\cos(Ax)\sin(By)-\cos\ll[A(\fr x-\fr\sqrt{3}y)\r]\sin\ll[B(\fr\sqrt{3} x+\fr y)\r]+\nn\\
\cos\ll[A(\fr x+\fr\sqrt{3}y)\r]\sin\ll[B(\fr\sqrt{3}x-\fr y)\r]\nn\\
B\neq 0
\label{30}
\ee
and
\be
\Psi_2^{as}(x,y,A,B)=\sin(Ax)\sin(By)+\sin\ll[A(\fr x-\fr\sqrt{3}y)\r]\sin\ll[B(\fr\sqrt{3} x+\fr y)\r]-\nn\\
\sin\ll[A(\fr x+\fr\sqrt{3}y)\r]\sin\ll[B(\fr\sqrt{3}x-\fr y)\r]\nn\\
A,B\neq 0
\label{31}
\ee

For the Neumann conditions on the parallelogram billiard boundary we have
\be
\Psi_1^{as}(x,y,A,B)=\cos(Ax)\cos(By)+\cos\ll[A(\fr x-\fr\sqrt{3}y)\r]\cos\ll[B(\fr\sqrt{3} x+\fr y)\r]+\nn\\
\cos\ll[A(\fr x+\fr\sqrt{3}y)\r]\cos\ll[B(\fr\sqrt{3}x-\fr y)\r]
\label{32}
\ee
and
\be
\Psi_2^{as}(x,y,A,B)=\sin(Ax)\cos(By)-\sin\ll[A(\fr x-\fr\sqrt{3}y)\r]\cos\ll[B(\fr\sqrt{3} x+\fr y)\r]-\nn\\
\sin\ll[A(\fr x+\fr\sqrt{3}y)\r]\cos\ll[B(\fr\sqrt{3}x-\fr y)\r]\nn\\
A\neq 0
\label{33}
\ee

It is to be noted as a general property of SWF's satisfying the Dirichlet boundary conditions that if one considers
points which are very close to the vertices of the parallelogram such as the one
shown in Fig.2.2 then differences between the phases of the component plane waves in the sum \mref{28} have to be small
just because of small distances between all emages of the original point and the point itself. One can conclude
therefore that values of the SWF's satisfying the Dirichlet boundary conditions should be the smaller at the points
mentioned the closer are the points to the vertices and of course have to vanish in the vertices. However a
sufficient distance of such points to the vertices depends
on a SWF considered and has to be the smaller the higher is the SWF energy eigenvalue, i.e. the smaller is the
corresponding wave length. In other words radii of circles centers of which coincide with the vertices and which contains
the points considered have to vanish if the momentum $p$ grows infinitely. As it was shown however by Hassel {\it et al}
\cite{39} vanishing of the circle areas enclosed by the vertices edges is such that the SWF square moduli integrated
over these areas are {\it finite} in the limit $p\to\infty$ while Marklof and Rudnick \cite{36} showed further that such
probabilities are the {\it same} in this limit as for the equal measure interior areas in polygons.

A similar general note can be done when one considers the respective behavior of SWF's satisfying the Neumannn
boundary conditions. However the corresponding conclusions seems to be quite opposite since in the formula
\mref{22} corresponding to the case the interference of the component plane waves is constructive for all points which
lie very close to the polygon vertices taking the value $2Ce^{\pm i(p_xx_k+p_yy_k)}$ in th $k^{th}$-vertex with
coordinates $(x_k,y_k)$. Because of that these SWF's cannot be normalized in polygons.

It is also worth to note that the set of the solutions \mref{30}-\mref{31} is just an example of incompleteness of the semiclassical states generated by
the method just applied. Namely putting $a=1$, i.e. reducing the parallelogram to the rhombus and shifting the origin
of the coordinate to the center of the latter and next rotating the axes to put the $x$-one on the longer diagonal of
the rhombus, one can check that both the solutions \mref{30}-\mref{31} are then odd under
the transformation $x\to -x$ while under the transformation $y\to -y$ the solution \mref{30} is even and \mref{31} is
odd. Therefore there are no among the solutions \mref{30}-\mref{31} the ones which are even under the reflection in the
$y$-axis \cite{8}.

\subsection{Single bay doubly rational broken rectangle billiards}

\hskip+2em Consider now a single bay broken rectangle billiard shown in Fig.3. Four independent periods are seen on the figure with the following
relations between them
\be
{\bf D}_{x2}=\frac{x_2}{x_1}{\bf D}_{x1}\nn\\
{\bf D}_{y2}=\frac{y_2}{y_1}{\bf D}_{y1}
\label{34}
\ee

Assuming the coefficients in \mref{34} to be all rational the billiard becomes doubly rational. It is obvious that such billiards are dens among
all billiards which are angle similar with the one in Fig.3. If $C_x,\;C_y$ are the least common multiples for the rational coefficients in \mref{34} we get
for the momenta and energy levels of the considered system
\be
p_{x,m}=\pi m\frac{C_x}{x_1}\nn\\
p_{y,n}=\pi n\frac{C_y}{y_1}\nn\\
E_{mn}=\fr\ll(p_{x,m}^2+p_{y,n}^2\r)=\fr\pi^2\ll(\frac{m^2C_x^2}{x_1^2}+\frac{n^2C_y^2}{y_1^2}\r)\nn\\
m,n=\pm 1,\pm 2,...
\label{35}
\ee

Considering SWF's corresponding to the spectra above we should note that there are four possible consistent sign prescriptions to the EPP from Fig.3
for which we get
\be
\Psi_{mn}^D(x,y)=\sin(p_{x,m}x)\sin(p_{y,n}y)=\sin\ll(\pi m\frac{C_x}{x_1}x\r)\sin\ll(\pi n\frac{C_y}{y_1}y\r)
\label{36}
\ee
satisfying the Dirichlet conditions on the billiard boundary and
\be
\Psi_{mn}^N(x,y)=\cos\ll(\pi m\frac{C_x}{x_1}x\r)\cos\ll(\pi n\frac{C_y}{y_1}y\r)
\label{37}
\ee
satisfying the Neumann conditions and according to the sign prescription shown in EPP of Fig.3
\be
\Psi_{mn}^{ND}(x,y)=\cos\ll(\pi m\frac{C_x}{x_1}x\r)\sin\ll(\pi n\frac{C_y}{y_1}y\r)
\label{38}
\ee
satisfying the Dirichlet conditions on the horizontal sides of the billiard and the Neumann ones on the vertical sides.

The last allowed consistent prescription exchanges the sin and cos functions in \mref{38} so that the  the Dirichlet conditions are satisfied
on the vertical sides of the billiard while the Neumann ones on the horizontal sides.

There are of course infinitely many global periodic skeletons one of which is horizontal with the period ${\bf D}_{x,1}/C_x$ and the other is vertical
with the period ${\bf D}_{y,1}/C_y$ . However since among the independent
periods of the considered billiard there are ones perpendicular to each other then quantizing on such skeletons we get the same energy
spectra and SWF's as for the aperiodic skeletons considered above. Although there is the high energy constraint \mref{21c} on quantizations on
such skeletons it is rather formal since in the case of DRPB's the obtained solutions are exact.

There are also infinitely many POC's the examples of which are those four ones on Fig.3 which are defined by the periods ${\bf D}_{x,i}$ and
${\bf D}_{y,i},\;i=1,2$.

\section{Incompleteness of the SWF approximations}

\hskip+2em This fact that the semiclassical conditions \mref{12a} have to be satisfied on every period direction is very demanding and limits
in fact applications of the SWF quantization to DRPB's. However despite the fact that in these cases the SWF quantization provides us with the {\it exact}
results it remains still an approximation in a sense that the energy spectra got by the method do not cover in general the whole spectra corresponding
to the cases considered. It means also that sets of SWF's accompanied the spectra are in general incomplete. The example of such a situation met in
the $\pi/3$-rhombus billiard was discussed at the end of sec.5.1 where all wave functions symmetric with respect to the longer axis could not be
constructed in the semiclassical wave function formalism presented in the paper.

Another aspects of losing states by the semiclassical quantization method used in our paper are provided by the broken rectangle billiards of Fig.3 for
which the set of the solutions \mref{36} satisfying the Dirichlet boundary condition cannot be considered as coinciding with the set of all solutions satisfying
this condition. In fact the set of the solutions \mref{36} follows as a subset of the {\it complete} ones vanishing on the boundary of the rectangle ABGF
if vanishing of the latter solutions on the lines $x=x_1$ and $y=y_1$ are additionally demanded.

To see that there are still more solutions which satisfy the Dirichlet boundary condition
in the broken rectangle ABCDEF which are not reconstructed by the SWF's \mref{36} consider two such rectangles with the following
sizes $x_1=1,\;x_2=2,\;y_1=1,\;y_2=2$ and $x_1=1,\;x_2=2-1/k,\;y_1=1,\;y_2=2$ where $k$ is natural and can be taken arbitrarily large. Quantizing the
latter we get according to \mref{36}
\be
\Psi_{mn}(x,y)=\sin(\pi mkx)\sin(\pi ny)\nn\\
E_{mk\;n}'=\pi^2(m^2k^2+n^2)\nn\\
m^2+n^2>0,\;m,n=0,\pm 1,\pm 2,...
\label{38a}
\ee
while for the former we get
\be
\Psi_{mn}(x,y)=\sin(\pi mx)\sin(\pi ny)\nn\\
E_{mn}=\pi^2(m^2+n^2)\nn\\
m^2+n^2>0,\;m,n=0,\pm 1,\pm 2,...
\label{38b}
\ee

It is seen from the above formulae that subsequent SWF's and the corresponding energy levels in \mref{38a} coincide only with every $k^{th}$ level of
the second spectrum for each fixed $n$.

On the other hand as it follows from {\bf THEOREM 1} of app.C both the spectra coincide up to $\epsilon$-accuracy with
$\epsilon=1/(k-1)$ for say $k>100$, i.e.
\be
\ll|\frac{E_{mn}'}{E_{mn}}-1\r|<\frac{1}{k-1}\nn\\
m^2+n^2>0,\;m,n=0,\pm 1,\pm 2,...
\label{38c}
\ee

Therefore most of the levels of the spectrum of the larger broken rectangle are absent in the smaller one despite
the fact that both the spectra have been obtained semiclassically.

Obviously a reason for that is a difference between the lengths of the largest
waves which can be used as the units for measuring the commensurate periods in the respective cases of the billiards. In the case of the smaller
broken rectangle
family the largest wave length is equal to $2/k$ while for the larger broken rectangle billiard it is equal to $2$, i.e. is $k$-times larger and it
cannot be used as a measuring unit for the smaller billiards as well as the shorter wave lengths equal to $2/r,\;r=2,3,...,k-1$, as the respective
subunits.

One can conclude therefore that the fuller is the description of energy spectra of DRPB's which the wave function semiclassical approximation provide us
with
the simpler are the fraction structures of the linear relations between the independent periods. The larger are both numerators and denominators in these
fractions the smaller have to be the largest wave lengths which can fit periods leading to still rare selections of the energy levels off their
total spectrum.

Therefore one can conclude that to get more energy levels for the cases of DRPB's with high fractions relating their periods
one should rather approximate these billiards by the ones for which these fractions are less complicated to get the respective energy levels as the
ones of the "simpler" billiards with some $\epsilon$-accuracy.

\section{Semiclassical quantization in any polygon billiard}

\hskip+2em A natural question which arises after the discussion of the semiclassical quantization in DRPB's done in the
previous sections is how the results got there can be extended to any polygon billiards, i.e. to the ones which are RPB's but not DRPB's as
well as to the irrational ones.

\begin{figure}
\begin{center}
\psfig{figure=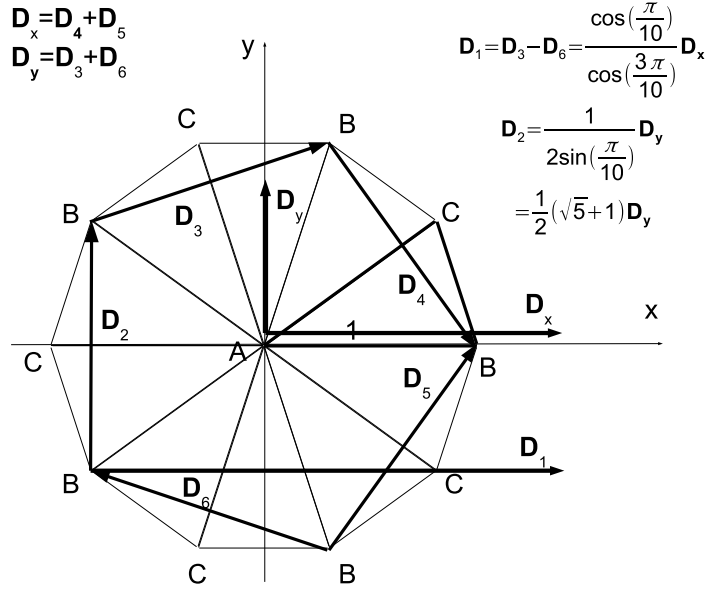,width=10cm} \caption{A symmetric EPP for the isosceles triangle ABC with the angle at the vertex A equal to $\pi/5$}
\end{center}
\end{figure}

Let us consider these questions consecutively.

\subsection{The RPB families containing the DRPB ones densely}

\hskip+2em In the case considered any RPB according to the {\bf THEOREM 2} of app.C can be approximated by DRPB's with any $\epsilon$-accuracy.
The spectra of DRPB's approximate then the
spectrum of the RPB considered with the respective $\epsilon$-accuracies. But according to our discussion in sec.6 one has to have in one's mind
that the better is the $\epsilon$-accuracy the more rare are sets of the approximated energy levels of the RPB and the more incomplete are the
sets of the respective SWF's.

\subsection{The RPB families deprived of dense DRPB subsets}

\hskip+2em A simple example of such a family is the one of the angle similar triangle billiards $ABC$ shown in Fig.5. Obviously up to a rescaling
all the member of the family are the same so that the irrational relations between the four independent periods ${\bf D}_1$ and ${\bf D}_x$ and
${\bf D}_2$ and ${\bf D}_y$ as shown in the figure are irremovable by any angle similar transformation of the triangle. Therefore the unique way
which allows us to apply the results of sec.6 to the case is to substitute the irrational numbers in these relations by their rational approximations
done with desired accuracy, i.e. by putting
\be
a_{ki}\approx\frac{p_{ki}}{q_{ki}}
\label{39}
\ee
for each such a coefficient where $p_{ki},q_{ki}$ are coprime integers with $q_{ki}>0$.

Once the substitutions \mref{39} are done the case considered becomes doubly rational and the procedure from the previous section can be applied
in its full extension. The unique question which arises when the respective results are got is how exact they are.

To get an answer to the question let us note that SWF's \mref{22} {\it cannot} be exact in this case since some of the considered periods were
approximated and these approximations cause that on the sides of the polygon which are the sides of the EPP used to construct the solutions \mref{22}
the latter do {\it not} satisfy the desired boundary conditions exactly. Nevertheless the conditions are certainly satisfied on nodal curves of the
SWF's which partly can coincide with the polygon boundary while the rest of them are very close to it. These nodal curves of SWF's are of course the
boundaries of domains for which the solutions are {\it exact}.

Let $D_{mn}$ be such a domain which corresponds to the solution $\Psi_{mn}^{as}$ and let
$D_P$ denote the domain occupied by the polygon considered. If $\frac{p_{ki}}{q_{ki}}\approx a_{ki}$ for all irrational $a_{ki}$ then we have to have
\be
D_{mn}\approx D_P
\label{40}
\ee
where the approximation in the above formula should be understood as measured by a number
$\epsilon_{mn}>0$ and is called the $\epsilon$-accuracy, see the {\bf DEFINITION 1} and the {\bf THEOREM 1} of app.C.

Let us enumerate the semiclassical energy levels $E_{mn}^{sem}$ by their growing values $E_{k_{mn}}^{sem},\;E_{k_{mn}}^{sem}\leq E_{k_{m'n'}}^{sem},
\;k_{mn}<k_{m'n'},\;k_{mn},k_{m'n'}=1,2,...$, and do the
same with the energy levels of the exact solutions.
Then we can arrange both the enumerations so that the property \mref{40} ensures according to the {\bf THEOREM 1} of app.C that for a given
$\epsilon_{k_{mn}}>0$ defining the $\epsilon$-accuracy in \mref{40} we have to have also
\be
\ll|\frac{E_{k_{mn}}^{sem}}{E_{Ck_{mn}}}-1\r|<\eta_{k_{mn}}
\label{41}
\ee
for all $k_{mn}$ where $\eta_{k_{mn}}$ depends on $\epsilon_{k_{mn}}$ and vanishes if does the latter and $C$ is a necessary rescaling coefficient discussed
earlier in sec.6.

\subsection{Semiclassical quantization in any of polygon billiards}

\hskip+2em Consider now an arbitrary polygon billiard, i.e. with irrational angles.
An obvious way which is in agreement with the spirit presented in the Introduction
would be a farther ''rationalization'' of the considered irrational polygon billiard (IPB) by approximating with desired accuracies
every of its irrational angles by rational ones, i.e. to substitute in this way any considered IPB by its respective
rational copy done with an arbitrary accuracy. With the RPB's obtained in this way we can proceed as it was described in the previous sections.
An accuracy of such an approach can be estimated again with the help of general theorems of the Courant and Hilbert monography \cite{40}, see also
app.C.

In practice however such an approach to the semiclassical quantization
of IPB's can mean using of the corresponding tori with very high genuses and therefore a necessity of establishing a huge number
of independent periods accompanying these tori.

Taking for example the right triangle with its remaining angles equal to $\frac{\sqrt{2}}{4}\pi$
and $\frac{2-\sqrt{2}}{4}\pi$ and approximating the latter angles by $0,353\pi$ and $0,147\pi$ respectively we should consider a corresponding
RPRS with $500$ independent periods and corresponding EPP's with
$2000$ component triangles, with the same number of sides corresponding to the multi torus containing $250$ holes.

Therefore a corresponding task seems to be very complicated and because of that discouraging
to the method. Nevertheless such an approach shows that at least theoretically there is a room for
quantizing semiclassically in the wave function formalism not only the pseudointegrable RPB's but also classically chaotic systems
which the IPB's are considered for.

\section{POC's, superscars and SWF's in global periodic skeletons}

\hskip+1.5em POC's in the global periodic skeletons are skeletons by themselves which in general are not global however.

POC's being not global periodic skeletons give rise however to the phenomenon known as superscars \cite{1,2} which has
been considered in our earlier paper \cite{38}.

A ray flow of each POC which exists in a global periodic skeleton is periodic under one of the independent periods of
the skeleton.
However its size perpendicular to its period depends on a geometry of the RPB considered being determined by
a distance between its two diagonals \cite{38}.

POC's being complete skeletons give rise for constructions on them both BSWF's as well as the corresponding
SWF's non-vanishing in the area of the polygon covered by a POC \cite{1,38}. In our earlier paper \cite{38} the constructed SWF's
satisfied the Dirichlet boundary
conditions in the polygon. However one can easily note that other boundary conditions are also allowed.

A good illustration of the latter statement are POC's shown in Fig.2.1. There are five POC's $P_i,\;i=1,...5$, with the
respective periods $\sqrt{3}(1+a),\;\sqrt{3},\;\sqrt{3}(1+a),\;\sqrt{3}(1+a),\;\sqrt{3}a$ and sides $\fr a,\;a-\fr,\;\fr a,\;\fr-\fr a,\;
1-\fr a$. On every of these POC's we can built SWF's periodic with respect to a corresponding period and satisfying {\it any} boundary
conditions on the POC sides. We can then use these POC SWF's to construct SWF's satisfying some boundary conditions in the parallelogram
billiard, i.e. to construct superscar solutions in the billiard \cite{1,38}.

Taking for example {\it any} solution mentioned above constructed on
the POC $P_1$ shown in Fig.2.4 and using the sign prescription shown in the third EPP of Fig.2 (note that this prescription is
{\it inconsistent} for building
a SWF for the global periodic skeleton shown in Fig.2.1) we can construct SWF's satisfying the Dirichlet and the Neumannn
conditions on the respective segments of the parallelogram billiard boundary shown in Fig.2.4 as well as any of these conditions on
the thin lines shown in Fig.2.4 being the $P_1$-POC diagonals.

This clearly shows that SWF's built in RPB's with the help
of respective SWF's constructed on POC's which can be found in such billiards have little to do with the SWF's constructed in sec.4 on the
global periodic skeletons or on the aperiodic ones.

In the paper mentioned \cite{38} the BSWF's build on POC's were periodic under the POC periods and vanished on their
diagonals. The
corresponding SWF's in the corresponding polygons were then built according to the rule given by \mref{22}. However the
corresponding sum in \mref{22} run {\it only} over
these image points of EPP's which laid inside {\it only} these emages of the basic polygon which were crossed by the
POC's. Therefore such SWF's vanished not only on the RPB boundaries but also on
the lines inside the polygons which were emages of the POC diagonals on which the normal derivatives of the SWF's were also
discontinuous.

In the context of the present paper SWF's and the corresponding energy spectra built in the above way in POC's crossing
a BRPB can be done periodic on the whole corresponding RPRS
just by shifting them by all possible {\it original} periods of the RPRS. In this way the POC SWF solutions become also
periodic on the RPRS with their spectra however having nothing
to do with the RPRS periods except of these single ones (and all their integer multiples) which are periods of POC's.

Comparing therefore SWF's built on POC's of a global periodic skeleton with the ones constructed on the skeleton itself
it is seen that if the latter is not a global single POC skeleton then
\begin{itemize}
\item SWF's built on POC's differ from the ones built on the global periodic skeleton itself;
\item SWF's built on a POC are periodic under the POC periods only;
\item energy spectra corresponding to POC's differ from the energy spectrum of the global periodic skeleton itself;
\item SWF's in POC's are {\it exact} and their set is {\it complete} in a rectangular domain occupied by a given POC .
\end{itemize}

In very rare cases of {\it global} single POC skeletons (met for example in the equilateral triangles or in the rectangles) the corresponding
SWF's and energy spectra coincide with the ones obtained by quantizing semiclassically the rational billiards mentioned on
aperiodic global skeletons. For such cases $k=0$ in the condition \mref{12d} leading to coincidences of the formulae
for the respective energy spectra.

Bogomolny and Schmit \cite{1,2} suggested that SWF's built on POC's of a global periodic skeleton can play some role in a
saturation of its quantum states. This suggestion has been however invalidated by Marklof and Rudnick \cite{36}. Also from
our discussion it follows rather just an opposite suggestion, i.e. that these are rather the POC states which can be expanded into the
SWF's built on the global skeletons.

In our earlier paper \cite{38} we have argued also that due to fact that POC's are perfect skeletons they can manifest
themselves as additional quantum states which can exist in the RPB's accompanying the billiard energy eigenstates. The
same POC's can be developed also in billiards which are completely different then the RPB's being also chaotic if they
only meet there
geometrical conditions allowing for their existence, i.e. they manifest themselves as a kind of resonant states in such
favourable conditions \cite{13,28,14,2}.

\section{Summary and discussion}

\hskip+2em In this paper we have demonstrated the method of the semiclassical quantization by the wave function approach of the
classically non integrable systems also chaotic ones represented by the polygon billiards. We have argued that it is in the spirit
of the wave function semiclassical quantization approach to rationalize respective quantities appearing in subsequent steps of
such an approach. This is due to the natural length measurers provided by lengths of waves naturally accompanying the
wave function formulation of the semiclassical approximation.

Let us enumerate the main steps which have been leading us to achieve our goal.
\begin{enumerate}
\item Construction of RPRS for RPB and revealing its periodic structure;
\item construction of EPP's for a RPB;
\item relating $2g$ independent periods of RPRS with a genus $g$ tori to which each EPP can be glued;
\item distinguishing DRPB's which when quantized semiclassically provide us with exact results for both the SWF's and energy spectra;
\item showing that the quantization of the classical billiard ball momenta are determined uniquely and only by periodic structure
of RPRS and BSWF's defined on it;
\item showing that BSWF's defined on any skeleton can have only the form of plane waves with classical or quantum momenta;
\item writing a general form of semiclassical energy spectra for any DRPB;
\item constructing SWF's with desired boundary conditions (Dirichlet, Neumannn or mixed ones) corresponding to given energy spectra and
satisfying the rules which such constructions are governed by due to necessity of satisfying by EPP's the consistent prescription constraints;
\item extending the semiclassical formalism built for DRPB's to RPB rationalizing linear relations between independent periods of RPB's
reducing this independence to only two of them chosen arbitrarily;
\item extending the semiclassical formalism built for RPB's to irrational ones by substituting the latter by RPB's approximating the
respective IPB with desired but otherwise arbitrary accuracies.
\end{enumerate}

The main conclusions which follow from the results of this paper are
\begin{itemize}
\item in principle any polygon billiards can be quantized semiclassically by the method presented in the paper;
\item semiclassical approximations obtained by the method are controlled by general theorems which can be found in \cite{40};
\item the semiclassical energy spectra described by the points 5. and 6. above are uniquely and only defined
for a given polygon by its respective system of independent periods;
\item there is a specific degeneration of energy spectra with respect to boundary conditions, i.e. there are many SWF's
corresponding to the same energy spectra but differing by boundary conditions they can satisfy;
\item there is no a full freedom in putting boundary conditions on SWF's in polygons so that
some semiclassical wave configurations and energy spectra corresponding to them can be ignored in the approach presented in this paper \cite{1,2};
\item there are specific conditions to be satisfied by periods of a RPRS (see formula \mref{12d}) which allow us for
constructions of SWF's on global periodic skeletons;
\item the wave function semiclassical quantization in the polygon billiards presented in the paper provides us with incomplete energy spectra and the
scale of this incompleteness is the larger the closer to the irrationality are the sizes of the polygon billiard angles.
\item in contrary to the previous conclusions it is always possible to construct respective SWF's, i.e. superscar solutions,
on any POC contained by global periodic skeletons which are exact and their set is complete in the rectangular domain occupied by a POC;
\item superscar solutions are resonant states in polygon and other billiards which exist in high energy regime due to
existence of classical POC's in such billiards and having rather little to do with the semiclassical limits of
eigenfunctions and energy spectra in quantum billiards \cite{38,36}
\end{itemize}

\appendix

\section{Polygon billiard skeleton dictionary}

\hskip+2em We have collected below the main notions used in the paper (see \cite{38,3} for their wider descriptions)
as well as the list of acronyms used in the paper frequently.

\begin{itemize}
\item{\bf rays} - classical trajectories in polygon billiards
\item{\bf reflections of rays} - reflections of rays by a side of a polygon billiard ruled by the reflection law of the geometrical optics
\item{\bf ray bundle (bundle)} - an open continuous set of rays parallel to each other starting from one side of a polygon billiard and ending on
another side
\item{\bf compound bundle} - a sum of two parallel bundles with a common boundary
\item{\bf periodic bundle} - a bundle containing only periodic trajectories with the same periods
\item{\bf global bundle} - a (compound) bundle which covers the whole polygon area
\item{\bf skeleton} - a set of bundles closed under reflections from sides of a polygon billiard
\item{\bf global skeleton} - a skeleton which each compound bundle is global
\item{\bf periodic skeleton} - a skeleton containing at least one periodic bundle
\item{\bf POC} - a single bundle periodic skeleton (periodic orbit channel \cite{1,2})
\item{\bf global aperiodic skeleton} - a global skeleton with no any periodic trajectory
\item{\bf global periodic skeleton} - a global skeleton with at least one POC
\item{\bf PB} - polygon billiard
\item{\bf RPB} - rational polygon billiard
\item{\bf BRPB} - basic rational polygon billiard
\item{\bf IPB} - irrational polygon billiard
\item{\bf RPRS} - rational polygon Riemann surface
\item{\bf ASP} - angle similar polygons
\item{\bf EPP} - elementary polygon pattern
\item{\bf DRPB} - doubly rational polygon billiard
\item{\bf SWF} - semiclassical wave function
\item{\bf BSWF} - basic semiclassical wave function
\end{itemize}

\section{BSWF's in global periodic skeletons}

\hskip+2em Consider a global periodic skeleton with $r-1$ POC's having respective periods ${\bf D}_k,\;k=2,...,r$ and
some number of aperiodic bundles assuming some local coordinate system. In the aperiodic bundles the corresponding BSWF's
have the same forms as in the aperiodic
global skeletons considered in sec.3.1, i.e. they are proportional to their exponential factors having therefore
the form
\be
\Psi_{k,n}^{ap}(x,y,p_n^{per})\equiv e^{\pm ip_n^{per}y}
\label{A20}
\ee
in the chosen local coordinate systems.

A BSWF $\Psi_{k,n}(x,y,p_n)$ however is defined on the $k^{th}$-POC by \mref{15}
and \mref{16}. Since $\chi_{k,0}^\sigma(x,y)$, $\sigma=\pm$, depend only on $x$, i.e.
$\chi_{k,0}^\sigma(x,y)\equiv \chi_{k,0}^\sigma(x)$ then \mref{16} for $k=0$ takes the following
form
\be
\chi_{k,1}^\sigma(x,y)=\chi_{k,1}^\sigma(x)+\frac{\sigma i}{2}y\ll((\chi_{k,0}^\sigma(x))''+2E_0\chi_{k,0}^\sigma(x)\r)
\label{A21}
\ee

Since further $\chi_{k,1}^\sigma(x,y)$ have to be periodic with respect to $y$ with periods ${\bf D}_k,\;k=2,...,r$ then
as it follows from \mref{A21} we have to have
\be
(\chi_{k,0}^\sigma(x))''+2E_{k,0}\chi_{k,0}^\sigma(x)=0
\label{A22}
\ee

To get a {\it periodic} solution of \mref{A22} we have to assume that $E_{k,0}>0$ so that
\be
\chi_{k,0}^\sigma(x)=C_{k,0}^+e^{i\sqrt{2E_{k,0}}x}+C_{k,0}^-e^{-i\sqrt{2E_{k,0}}x}
\label{A23}
\ee
and additionally we get the independence of $\chi_{k,1}^\sigma(x,y)$ on $y$, i.e. $\chi_{k,1}^\sigma(x,y)\equiv\chi_{k,1}^\sigma(x)$.

Next we have to enforce on $\chi_{k,0}^\sigma(x)$ its periodicity on ${\bf D}_1/C_1$, i.e.
\be
\chi_{k,0}^\sigma(x)=\chi_{k,0}^\sigma(x+D_{1x}/C_1)=\chi_{k,0}^\sigma(x+D_1\sin\alpha/C_1)
\label{A24}
\ee
where $\alpha$ is the angle between the periods ${\bf D}_1$ and ${\bf D}_2$.

Consequently we have to have further
\be
\sqrt{2E_{k,0}}D_1\sin\alpha=2m\pi C_1\;\;\;\;\;\;\;k=2,...,r,\;m=1,2,...,
\label{A25}
\ee

Making the next step in solving \mref{16} we get for $\chi_{k,1}^\sigma(x)$ the following equation
\be
(\chi_{k,1}^\sigma(x))''+2E_{k,0}\chi_{k,1}^\sigma(x)=-2E_{k,1}\chi_{k,0}^\sigma(x)
\label{A26}
\ee
with the solution
\be
\chi_{k,1}^\sigma(x)=C_{k,1}^+e^{i\sqrt{2E_{k,0}}x}+C_{k,1}^-e^{-i\sqrt{2E_{k,0}}x}-
\frac{E_{k,1}}{E_{k,0}}x(\chi_{k,0}^\sigma(x))'
\label{A27}
\ee

Again the periodicity of $\chi_{k,1}^\sigma(x)$ on ${\bf D}_1/C_1$ demands $E_{k,1}=0$. By induction we then get
$E_{k,l}=0,\;l\geq 1$, and finally
\be
\chi_k^\sigma(x,y,p_n)\equiv C_k^+(p_n)e^{i\sqrt{2E_{k,0}}x}+C_k^-(p_n)e^{-i\sqrt{2E_{k,0}}x}\nn\\
C_k^\pm(p_n)=\sum_{l\geq 0}\frac{C_{k,l}^\pm}{p_n^l}
\label{A28}
\ee

\section{Smooth behavior of energy levels as a function of a billiard boundary - a general theorem and an example}

\hskip+2em Consider two single bay broken rectangles shown in Fig.3 and having the bottom side lengths equal to $x_2$ and $x_3$. If the latter is
arbitrarily close to $x_2$ then energy spectra of both the
billiards are also arbitrarily close to each other in the sens of the following theorem proved in the monography of Courant and Hilbert
\cite{40}, p.421.

\begin{de} It is said that the domain $G$ is approximated by the domain $G'$ with the $\epsilon$-accuracy if $G$ together with its
boundary can be transformed pointwise into the domain $G'$ together with its boundary by the equations
\be
x'=x+g(x,y)\nn\\
y'=y+h(x,y)
\label{A29}
\ee
where $g(x,y),\;h(x,y)$ are both piecewise continuous and less in $G$ in their absolute values than a small positive number $\epsilon$ together with
their first derivatives.
\end{de}

\begin{de}
If all conditions of {\bf DEFINITION 1} are satisfied while $\epsilon\to 0$ then it is said that $G$ is a continuous deformation of $G'$.
\end{de}

\begin{tw}
Let $G$ and $G'$ satisfy all conditions of the {\bf DEFINITION 1}. Then for any boundary condition $\p\Psi/\p n+\sigma\Psi=0$ the energy spectrum
corresponding to $G'$ approximates the one of $G$ with the $\epsilon$-accuracy. More precisely for any $\epsilon$ there is a number $\eta$
depending only on $\epsilon$ and vanishing with it such that for respectively ordered energy levels $E'_n$ and $E_n$ corresponding to the domains
$G'$ and $G$ we have
\be
\ll|\frac{E_n'}{E_n}-1\r|<\eta
\label{A30}
\ee
\end{tw}

\begin{tw}
Let $G$ and $G'$  satisfy the conditions of the {\bf THEOREM 1} and $G$ is a continuous deformation of $G'$ then the energy spectrum
corresponding to $G'$ varies continuously with $\epsilon\to 0$ approaching the energy spectrum of $G$ controlled by the conditions \mref{A30}.
\end{tw}

Applying the {\bf THEOREM 2} to the case of Fig.3 where $G$ coincides with the broken rectangle $ABCDEF$ while $G'$ with the $AB'C'DEF$ one
it is enough to construct the respective transformations \mref{A29}. This can be done as follows
\be
\ba{l}x'=\ll\{\ba{lr}x&0\leq x\leq x_1,\;0\leq y\leq y_2\\
              x-\frac{x-x_1}{x_2-x_1}(x_2-x_3)&x_1\leq x\leq x_2,\;0\leq y\leq y_1
              \ea\r.\\
      y'=y
\ea
\label{A31}
\ee
with $\epsilon=x_2-x_3$ if $x_2-x_1>1$ and with $\epsilon=\frac{x_2-x_3}{x_2-x_1}$ in the opposite case assuming that $x_2-x_3<<x_2-x_1$.

\end{document}